\documentclass[aps,pre,twocolumn,groupedaddress]{revtex4-2}
\usepackage{amsmath}
\usepackage[dvipdfmx]{graphicx}
\usepackage{color}
\usepackage{comment}
\usepackage{lineno}

\begin{document}


\title{Achieving Carnot efficiency in a finite-power Brownian Carnot cycle with arbitrary temperature difference}


\author{Kosuke Miura, Yuki Izumida,$^{\dagger}$ and Koji Okuda}
\affiliation{Department of Physics, Hokkaido University, Sapporo 060-0810, Japan\\
  $^{\dagger}$Department of Complexity Science and Engineering, Graduate School of Frontier Sciences, The University of Tokyo, Kashiwa 277-8561, Japan}


\date{\today}

\begin{abstract}
Achieving the Carnot efficiency at finite power is a challenging problem in heat engines due to the trade-off relation between efficiency and power that holds for general heat engines. 
It is pointed out that the Carnot efficiency at finite power may be achievable in the vanishing limit of the relaxation times of a system without breaking the trade-off relation.
However, any explicit model of heat engines that realizes this scenario for arbitrary temperature difference has not been proposed. 
Here, we investigate an underdamped Brownian Carnot cycle where the finite-time adiabatic processes connecting the isothermal processes are tactically adopted. 
We show that in the vanishing limit of the relaxation times in the above cycle, the compatibility of the Carnot efficiency and finite power is achievable for arbitrary temperature difference. 
This is theoretically explained based on the trade-off relation derived for our cycle, which is also confirmed by numerical simulations.

\end{abstract}


\maketitle

\section{Introduction}
In recent years, understanding the relation between efficiency and power in heat engines has been recognized as an important issue in nonequilibrium thermodynamics.
The Carnot efficiency $\eta_{\rm C}$ gives the upper bound of the efficiency $\eta$ of any heat engine operating between hot and cold heat baths with the temperature $T_h$ and $T_c (<T_h)$,
\begin{equation}
\label{Carnot efficiency}
	 \eta  \le \eta_{\rm C} \equiv 1-\frac{T_c}{T_h},
\end{equation}
where $\eta$ is defined as the ratio of the work to the heat from the hot heat bath~\cite{Carnot_book,Callen_2nd_book}.
The Carnot efficiency can usually be realized in the quasistatic Carnot cycle that takes an infinitely long cycle time.
However, the power $P$ defined as the output work per unit time vanishes in the quasistatic limit.
Since the compatibility of the Carnot efficiency and finite power may not be forbidden by the second law of thermodynamics, many efforts have been made to achieve it~\cite{PhysRevLett.106.230602,Polettini_2017,PhysRevE.62.6021,PhysRevE.95.052128,
  PhysRevE.62.6021,Campisi2016,
  PhysRevLett.110.070603,PhysRevB.87.165419,PhysRevB.94.121402,
  PhysRevLett.112.140601,PhysRevLett.114.146801,Sothmann_2014,PhysRevE.98.042112,PhysRevLett.124.110606,PhysRevE.96.062107,PhysRevLett.121.120601,PhysRevE.103.042125,Lee_2017,PhysRevE.101.052132}.
This possibility, however, may be reconsidered in terms of the recently formulated trade-off relation between the efficiency and power in heat engines~\cite{PhysRevLett.117.190601,PhysRevLett.120.190602,Koyuk_2018,Dechant_2018,PhysRevE.97.062101},
\begin{equation}
\label{generalized trade-off relation}
	P\leq A\eta(\eta_{\rm C}-\eta),
\end{equation}
where $A$ is a positive quantity depending on the system.
Applicability of the trade-off relation ranges from macroscopic heat engines~\cite{PhysRevLett.117.190601} to stochastic ones~\cite{PhysRevLett.120.190602,Koyuk_2018,Dechant_2018,PhysRevE.97.062101} described by the Markovian dynamics.
According to this relation, as the efficiency $\eta$ approaches $\eta_{\rm C}$, the power vanishes, which means that the compatibility of the Carnot efficiency and finite power is forbidden.

On the other hand, if the quantity $A$ in Eq.~(\ref{generalized trade-off relation}) diverges at the same time as $\eta$ approaches $\eta_{\rm C}$, the power may remain finite.
By focusing on the dependence of $A$ on the relaxation times, Holubec and Ryabov proposed a scenario to achieve the Carnot efficiency at finite power without breaking the trade-off relation by taking the vanishing limit of the relaxation times, which can yield a diverging quantity $A$~\cite{PhysRevLett.121.120601}.
However, complicated dependence of the quantity $A$ and the efficiency $\eta$ on the relaxation times may make the feasibility of the scenario nontrivial.

In Ref.~\cite{PhysRevE.103.042125}, we studied the Brownian Carnot cycle to demonstrate the feasibility of the above scenario.
We used an underdamped Brownian Carnot cycle, where the instantaneous change of the potential and temperature of the heat bath is used as an adiabatic process~\cite{PhysRevLett.121.120601,Dechant_2017,Schmiedl_2007,PhysRevE.97.022131}.
We numerically and theoretically showed the compatibility of the Carnot efficiency and finite power in the vanishing limit of the relaxation times in the small temperature-difference regime.
This result was obtained by explicitly expressing the relaxation-times dependence of $A$ and $\eta$ in Eq.~(\ref{generalized trade-off relation}).
In the large temperature-difference regime, however, we cannot achieve the compatibility even in the vanishing limit of the relaxation times.
Just after the instantaneous adiabatic processes, there exists a relaxation of the kinetic energy of the particle.
The heat flowing in the relaxation is regarded as the inevitable heat leakage that reduces the efficiency. 
Since the heat leakage is proportional to the temperature difference, we cannot neglect it for a large temperature difference.
Thus, the compatibility of the Carnot efficiency and finite power has not been established yet without the restriction of the small temperature difference.

In this paper, we will show that the compatibility of the Carnot efficiency and finite power is achievable in the vanishing limit of the relaxation times in an underdamped Brownian Carnot cycle with arbitrary temperature difference, where finite-time adiabatic processes~\cite{PhysRevLett.114.120601,PhysRevE.101.032129,Plata_2020} are introduced instead of the instantaneous ones.
To be more specific, we consider a spatially one-dimensional system and assume that in this cycle, the Brownian particle is in contact with the heat bath with time-dependent temperature $T(t)$ and trapped by a harmonic potential,
\begin{equation}
\label{harmonic potential}
	V(x,t)=\frac{1}{2}\lambda(t)x^2,
\end{equation}
where the stiffness $\lambda(t)$ depends on the time $t$.
Then, a finite-time adiabatic process can be realized by carefully controlling both the temperature of the heat bath and stiffness to prevent the heat flowing at any time during the process.
Here, note that the word ``finite" in this paper means ``nonzero" and ``noninfinite."
For example, the finite-time adiabatic process means the adiabatic process where the time taken for this process is not zero and not infinite.
Remarkably, this carefully controlled adiabatic process can eliminate the heat leakages that exist in the instantaneous adiabatic process because of the continuous nature of the process.
From detailed calculations, we can show that $A$ in Eq.~(\ref{generalized trade-off relation}) in our cycle diverges while making the entropy production per cycle vanish, under the fixed cycle time in the vanishing limit of the relaxation times for arbitrary temperature difference.
Therefore, we can establish the compatibility of the Carnot efficiency and finite power within the framework of the trade-off relation in Eq.~(\ref{generalized trade-off relation}).

This paper is organized as follows.
In Sec.~\ref{model}, we introduce the probability distribution of the Brownian particle trapped by the harmonic potential and describe its time evolution by the Kramers equation.
We also introduce the thermodynamic processes, where the Brownian particle is in contact with the heat bath at any time, and consider them in the small relaxation-times regime.
In Sec.~\ref{finite-time adiabatic process}, we consider the finite-time adiabatic processes.
In Sec.~\ref{Carnot cycle in the small relaxation-times regime}, we construct the Carnot cycle by using the isothermal and finite-time adiabatic processes.
In Sec.~\ref{theoretical analysis}, we theoretically show that the compatibility of the Carnot efficiency and finite power is achievable without breaking the trade-off relation in Eq.~(\ref{generalized trade-off relation}).
In Sec.~\ref{numerical simulation}, we present the results of numerical simulations of our cycle when we vary the temperature difference and relaxation times of the system.
In Sec.~\ref{summary and discussion}, we summarize this paper.

\section{model}
\label{model}
We consider the underdamped Brownian particle in contact with the heat bath with the time-dependent temperature $T(t)$ and trapped by the harmonic potential in Eq.~(\ref{harmonic potential}).
We can describe the dynamics of the Brownian particle by the underdamped Langevin equations given by
\begin{align}
  \label{Langevin equation of x}
  \dot{x} =& v,\\
  \label{Langevin equation of v}
  m \dot{v} =& -\gamma v - \lambda x +\sqrt{2\gamma k_{\rm B} T} \xi,
\end{align}
where $x$, $v$, and $m$ are the position,
velocity, and mass of the particle, respectively.
In the following, we set the Boltzmann constant $k_{\rm B}=1$ for simplicity.
$\gamma$ is the friction constant and is assumed to be independent of the temperature $T(t)$.
The Gaussian white noise $\xi(t)$ satisfies $\langle \xi(t)\rangle = 0$ and $\langle \xi(t)\xi(t')\rangle = \delta(t-t')$, where $\langle \cdots \rangle$ denotes the statistical average.
The dot denotes the time derivative or a quantity per unit time.
Then, we introduce the probability distribution $p(x,v,t)$ to describe the state of the Brownian particle.
Its time evolution is described by the Kramers equation~\cite{Risken_2nd_book},  given by
\begin{align}
  \label{Fokker-Planck equation}
    \frac{\partial}{\partial t}&p(x,v,t)
     = -\frac{\partial}{\partial x}(v p(x,v,t))\nonumber\\
    &+ \frac{\partial}{\partial v}\left[
       \frac{\gamma}{m}v + \frac{\lambda}{m}x
       + \frac{\gamma T}{m^2}\frac{\partial}{\partial v}
       \right]p(x,v,t).
\end{align}
The relaxation times of position $\tau_{x}$ and velocity $\tau_{v}$ of the Brownian particle are defined as
\begin{align}
  \label{relaxation time of x}
  \tau_{x}(t) \equiv & \frac{\gamma}{\lambda(t)},\\
  \label{relaxation time of v}
  \tau_{v} \equiv &\frac{m}{\gamma},
\end{align}
where $\tau_{x}(t)$ depends on time through the stiffness $\lambda(t)$.
We fix $\gamma$ and change $\tau_{x}$ and $\tau_v$ by changing $m$ and $\lambda$.
$\tau_x$ and $\tau_v$ denote the time constants that determine the speed of relaxation of the position and velocity of the particle, respectively, into their equilibrium values.
Here, we define $\sigma_{x}(t) \equiv \langle x^2 \rangle$, $\sigma_{v}(t) \equiv \langle v^2 \rangle$, and $\sigma_{xv}(t) \equiv \langle xv \rangle$.
Then, assuming that the probability distribution $p(x,v,t)$ is Gaussian, we obtain
\begin{equation}
  \label{Gaussian distribution}
    p(x,v,t) = \frac{1}{\sqrt{4 \pi^2\Phi}}\exp \left\{ -\frac{\sigma_{x}v^2 + \sigma_{v}x^2 - 2\sigma_{xv} xv}{2\Phi}\right\},
\end{equation}
where we defined the quantity $\Phi(t)$ as
\begin{equation}
\label{Phi}
	\Phi(t)\equiv \sigma_x(t)\sigma_v(t)-\sigma_{xv}(t)^2.
\end{equation}
From the Cauchy-Schwarz inequality, $\Phi$ should satisfy
\begin{equation}
\label{inequality of the variables}
	\Phi(t)\geq 0.
\end{equation}
Then, from Eqs.~(\ref{Gaussian distribution}) and (\ref{Phi}), we can see that the state of the Brownian particle is described by only the three variables $\sigma_{x}(t)$, $\sigma_{v}(t)$, and $\sigma_{xv}(t)$.
From Eq.~(\ref{Fokker-Planck equation}), we can derive the time evolution equations of $\sigma_{x}$, $\sigma_{v}$, and $\sigma_{xv}$~\cite{Dechant_2017} as
\begin{align}
  \label{equation of sigma_x}
  &\dot{\sigma}_{x} = 2\sigma_{xv},\\
  \label{equation of sigma_v}
  &\dot{\sigma}_{v} = \frac{2 \gamma T}{m^2} - \frac{2 \gamma}{m}\sigma_{v}
  - \frac{2 \lambda}{m}\sigma_{xv},\\
  \label{equation of sigma_xv}
  &\dot{\sigma}_{xv} = \sigma_{v} - \frac{\lambda}{m}\sigma_{x} - \frac{ \gamma}{m}\sigma_{xv}.
\end{align}
We can use Eqs.~(\ref{equation of sigma_x})--(\ref{equation of sigma_xv}) to describe the time evolution of the system instead of Eq.~(\ref{Fokker-Planck equation}).
Under the Gaussian distribution in Eq.~(\ref{Gaussian distribution}), the internal energy $E(t)$ and entropy $S(t)$ of the particle are given by
\begin{align}
  \label{def of the internal energy}
    E(t) &\equiv \int^{\infty}_{-\infty} dx \int^{\infty}_{-\infty} dv\ p(x,v,t)\ \left[ \frac{1}{2}mv^2 + \frac{1}{2}\lambda(t) x^2\right]\nonumber\\
    &= \frac{1}{2}m\sigma_{v}(t) + \frac{1}{2}\lambda(t) \sigma_{x}(t),
\end{align}
\begin{align}
  \label{def of the entropy}
    S(t) \equiv& -\int^{\infty}_{-\infty} dx\int^{\infty}_{-\infty} dv\ p(x,v,t)\ln\{p(x,v,t)\}\nonumber\\
    =&\frac{1}{2}\ln\Phi(t)+\ln(2\pi)+1.
\end{align}

\subsection{Thermodynamic process}
\label{Thermodynamic process}
We consider a thermodynamic process lasting for $t_i\leq t \leq t_f$.
We define $X_i\equiv X(t_i)$, $X_f\equiv X(t_f)$, and 
\begin{equation}
\label{def of Delta}
    \Delta X\equiv X_f-X_i
\end{equation}
for any physical quantity $X(t)$.
Then, the time taken for this process is given by
\begin{equation}
\label{def of time interval}
    \Delta t\equiv t_f-t_i.
\end{equation}
In the thermodynamic process, we operate $T(t)$ and $\lambda(t)$.
Generally, they are functions of $t$, including their initial and final values and $\Delta t$.
We call these functions a protocol.
We also call the thermodynamic process in contact with the heat bath with the constant temperature the isothermal process,
where we can choose $\lambda(t)$ independently of the constant $T$.
On the other hand, in the finite-time adiabatic process, $T(t)$ and $\lambda(t)$ cannot be chosen independently once $\Delta t$ is fixed (see Sec.~\ref{finite-time adiabatic process}).

To define the work and heat, we consider the internal-energy change rate given by
\begin{align}
\label{internal energy change}
    \dot{E}(t)=& \int^{\infty}_{-\infty} dx\int^{\infty}_{-\infty} dv\ \frac{\partial p(x,v,t)}{\partial t}\left[ \frac{1}{2}mv^2 + \frac{1}{2}\lambda(t) x^2\right]\nonumber\\
    &\ +\int^{\infty}_{-\infty} dx\int^{\infty}_{-\infty} dv\ p(x,v,t)\ \left(\frac{1}{2}\dot{\lambda}(t)x^2\right),
\end{align}
using Eq.~(\ref{def of the internal energy}).
The second term on the right-hand side of Eq.~(\ref{internal energy change}) represents the internal-energy change rate caused by the change of $\lambda(t)$.
We can regard this energy change rate as the work per unit time done to the Brownian particle.
Thus, we define the output work per unit time as the negative value of the second term of Eq.~(\ref{internal energy change}), which is given by
\begin{align}
  \label{def of the work per unit time}
	\dot{W}(t) \equiv& -\int^{\infty}_{-\infty} dx\int^{\infty}_{-\infty} dv\ p(x,v,t)\ \left(\frac{1}{2}\dot{\lambda}(t)x^2\right)\nonumber\\
	=&- \frac{1}{2}\dot{\lambda}(t) \sigma_{x}(t).
\end{align} 
From the first law of thermodynamics and $\dot{W}$ in Eq.~(\ref{def of the work per unit time}), we can define the heat flux $\dot{Q}$ flowing from the heat bath to the particle as
\begin{align}
\label{def of the heat flux}
	\dot{Q}(t)\equiv& \dot{W}(t)+\dot{E}(t)\nonumber\\
	=&\frac{1}{2}m\dot{\sigma}_{v}(t)+\frac{1}{2}\lambda(t)\dot{\sigma}_{x}(t)\nonumber\\
	=&\frac{\gamma}{m}[T(t)-m\sigma_v(t)],
\end{align}
using Eqs.~(\ref{equation of sigma_x}), (\ref{equation of sigma_v}), (\ref{internal energy change}), and (\ref{def of the work per unit time}).
Then, from Eqs.~(\ref{def of the work per unit time}) and (\ref{def of the heat flux}), we obtain the output work $W$ and heat $Q$ in this process as
\begin{align}
    \label{def of the work}
    W \equiv& -\int^{t_f}_{t_i}dt\ \frac{1}{2}\dot{\lambda}(t) \sigma_{x}(t),\\
    \label{def of the heat}
    Q\equiv&\int^{t_f}_{t_i}dt\ \frac{\gamma}{m}[T(t)-m\sigma_v(t)] = W+\Delta E.
\end{align}
Using Eqs.~(\ref{equation of sigma_x})--(\ref{equation of sigma_xv}), (\ref{def of the entropy}), and (\ref{def of the heat flux}), we can derive the entropy production rate $\dot{\Sigma}$ of the total system including the particle and heat bath as 
\begin{align}
\label{def of the entropy production rate}
	\dot{\Sigma}(t) \equiv& \dot{S}(t) - \frac{\dot{Q}(t)}{T(t)}\nonumber\\
	=&\frac{\frac{\gamma}{m}\left(T-m\sigma_v\right)^2+ (2T-m\sigma_{v})\gamma\frac{\sigma_{xv}{}^2}{\sigma_{x}}}{T\left(m\sigma_{v}-m\frac{\sigma_{xv}{}^2}{\sigma_{x}}\right)}\nonumber\\	
	=&\frac{m}{\gamma}\frac{\dot{Q}^2}{Tm\sigma_v}+\frac{\gamma}{m}
 	\frac{T\sigma_{xv}{}^2}{m\sigma_v\left(\sigma_{x}\sigma_{v}-\sigma_{xv}{}^2\right)} \geq 0,
\end{align}
where the last inequality comes from Eq.~(\ref{inequality of the variables}).
From Eq.~(\ref{def of the entropy production rate}), we obtain the entropy production of the total system in this process as
\begin{align}
\label{def of the entropy production}
	\Sigma =& \int^{t_f}_{t_i}dt\ \dot{\Sigma}(t)\nonumber\\
	= & \Delta S -\int^{t_f}_{t_i}dt\ \frac{\dot{Q}(t)}{T(t)},
\end{align}
where $\Delta S \equiv S_f-S_i$ is the entropy change of the particle in this process.

\subsection{Thermodynamic process in the small relaxation-times regime}
\label{thermodynamic processes in the small relaxation-times regime}
We consider the thermodynamic process mentioned in Sec.~\ref{Thermodynamic process}, in the small relaxation-times regime where the relaxation times in Eqs.~(\ref{relaxation time of x}) and (\ref{relaxation time of v}) are sufficiently smaller than $\Delta t=t_f-t_i$ at any time.
Below, we refer to ``small relaxation time" when $\tau_j/\Delta t \ll 1$ ($j=x,v$) is satisfied.
We assume that $\Delta t$ is finite in this regime.
For convenience, we introduce a normalized time:
\begin{equation}
\label{normalized time}
    s\equiv \frac{t-t_i}{\Delta t}\quad (0\leq s\leq 1).
\end{equation}
In the small relaxation-times regime in this process, we assume that $T(t)$ and $\lambda(t)$ ($t_i\leq t\leq t_f$) depend on time and vary smoothly and slowly, that is, on a time-scale sufficiently longer than the relaxation times $\tau_x$ and $\tau_v$.
Moreover, we also assume that $T(t')/T(t)$ and $\lambda(t')/\lambda(t)$ are finite at any times $t$ and $t'$ ($t_i\leq t'\leq t_f$).
We can rewrite $\lambda(t')/\lambda(t)$ as
\begin{equation}
\label{ratio of the relaxation times of x}
    \frac{\lambda(t')}{\lambda(t)}=\frac{\tau_x(t)}{\tau_x(t')},
\end{equation}
by using Eq.~(\ref{relaxation time of x}).
From Eq.~(\ref{ratio of the relaxation times of x}), we find that the assumption above Eq.~(\ref{ratio of the relaxation times of x}) means that when we make $\tau_x(t)$ small, $\tau_x(t')$ for any $t'$ in the same process also becomes small.
From the Taylor expansion, we have
\begin{align}
    \label{Taylor expansion of temperature}
    \frac{T(t+\delta t)}{T(t)} \simeq 1+\delta t \frac{d}{dt}\ln T(t),\\
    \label{Taylor expansion of stiffness}
    \frac{\lambda(t+\delta t)}{\lambda(t)} \simeq 1+\delta t \frac{d}{dt}\ln\lambda(t),
\end{align}
where $\delta t$ is assumed to be sufficiently smaller than $\Delta t$.
Since, from the assumption above Eq.~(\ref{ratio of the relaxation times of x}), the left-hand side of Eqs.~(\ref{Taylor expansion of temperature}) and (\ref{Taylor expansion of stiffness}) are finite, $d(\ln T)/dt$ and $d(\ln \lambda)/dt$ do not diverge even when the relaxation time of position vanishes, in other words, $\lambda(t)$ in Eq.~(\ref{relaxation time of x}) diverges.
As shown in Appendix \ref{Derivation of Eq ref (approximation of variables)}, the variables $\sigma_x$, $\sigma_v$, and $\sigma_{xv}$ in the small relaxation-times regime can be approximated by
\begin{equation}
\label{approximation of variables}
	\sigma_x\simeq\frac{T}{\lambda}, \quad \sigma_v\simeq\frac{T}{m}, \quad \sigma_{xv}\simeq\frac{1}{2}\frac{d}{dt}\left(\frac{T}{\lambda}\right).
\end{equation}
We can rewrite $\sigma_{xv}$ in Eq.~(\ref{approximation of variables}) as
\begin{equation}
\label{sigma xv in the small relaxation times regime}
    \sigma_{xv}\simeq\frac{T}{2\lambda}\frac{d}{dt}\ln\left(\frac{T}{\lambda}\right)=\frac{\tau_x T}{2\gamma}\frac{d}{dt}\ln\left(\frac{T}{\lambda}\right),
\end{equation}
using Eq.~(\ref{relaxation time of x}).
Because $d(\ln T)/dt$ and $d(\ln \lambda)/dt$ do not diverge, $\sigma_{xv}$ vanishes when $\tau_x$ vanishes.
Then, since we can neglect $\sigma_{xv}{}^2$ compared with $\sigma_x\sigma_v$, $\Phi$ in Eq.~(\ref{Phi}) is approximated by
\begin{equation}
    \label{approximation of Phi}
    \Phi\simeq \frac{T^2}{m\lambda}.
\end{equation}

We consider the heat flux in Eq.~(\ref{def of the heat flux}) in the small relaxation-times regime.
In the vanishing limit of the relaxation times, $\dot{Q}$ in Eq.~(\ref{def of the heat flux}) appears to diverge since $\gamma/m=1/\tau_v$ diverges, which is the coefficient of $T-m\sigma_v$ in Eq.~(\ref{def of the heat flux}).
However, since $m\sigma_v$ approaches $T$ in the above limit from Eq.~(\ref{approximation of variables}), $\dot{Q}$ may become finite.
In fact, we can evaluate $\dot{Q}$ using Eq.~(\ref{approximation of variables}) and the second line of Eq.~(\ref{def of the heat flux}).
By using Eq.~(\ref{time dependence of sigma}), we obtain
\begin{equation}
  \label{time dependence of sigma x and sigma v}
  \lambda(t)\dot{\sigma}_{x}(t)\simeq T\left(\frac{d}{dt}\ln \frac{T}{\lambda}\right),
  \quad
  m\dot{\sigma}_{v} \simeq \dot{T}=T\frac{d}{dt}\ln T.
\end{equation}
Then, the heat flux $\dot{Q}(t)$ in Eq.~(\ref{def of the heat flux}) is approximated by
\begin{equation}
  \label{heat flux with small relaxation times}
    \dot{Q}(t) =\frac{1}{2}m\dot{\sigma}_{v}+\frac{1}{2}\lambda\dot{\sigma}_{x}
    \simeq \frac{T}{2}\left(\frac{d}{dt}\ln \frac{T^2}{\lambda}\right).
\end{equation}
Since $d(\ln T)/dt$ and $d(\ln \lambda)/dt$ do not diverge even in the vanishing limit of the relaxation times, $\dot{Q}$ does not diverge.
Similarly, the entropy production rate in Eq.~(\ref{def of the entropy production rate}) is approximated by
\begin{align}
\label{entropy production rate in the small relaxation-times regime}
\dot{\Sigma}(t)\simeq& \frac{1}{T}\frac{\tau_{v}\dot{Q}^2
    + \frac{\tau_x T^2}{4}\left(\frac{d}{dt}\ln\frac{T}{\lambda}\right)^2}
  {T-\frac{\tau_v\tau_x T}{4}\left(\frac{d}{dt}\ln\frac{T}{\lambda}\right)^2}\nonumber\\
  =&\frac{1}{\Delta t}\frac{1}{T}\frac{\frac{\tau_{v}}{\Delta t}\left(\frac{dQ}{ds}\right)^2
    + \frac{\tau_x T^2}{4{\Delta t}}\left(\frac{d}{ds}\ln\frac{T}{\lambda}\right)^2}
  {T-\frac{\tau_v}{\Delta t}\frac{\tau_x}{\Delta t}\frac{T}{4}\left(\frac{d}{ds}\ln\frac{T}{\lambda}\right)^2},
\end{align}
where we used Eqs.~(\ref{relaxation time of x}), (\ref{relaxation time of v}), (\ref{normalized time}), and (\ref{approximation of variables}).
Since $\dot{Q}$ and $d\ln(T/\lambda)/dt$ do not diverge, and $\Delta t$ is finite, we can see that $dQ/ds$ and $d\ln(T/\lambda)/ds$ do not diverge.
Thus, we can see that the entropy production rate $\dot{\Sigma}$ vanishes in the vanishing limit of $\tau_x$ and $\tau_v$.
The entropy production in this process in the small relaxation-times regime is given by
\begin{align}
\label{entropy production in the small relaxation-times regime}
	\Sigma\simeq&\int^{1}_{0}ds \frac{1}{T}\frac{\frac{\tau_{v}}{\Delta t}\left(\frac{dQ}{ds}\right)^2
    + \frac{\tau_x T^2}{4{\Delta t}}\left(\frac{d}{ds}\ln\frac{T}{\lambda}\right)^2}
  {T-\frac{\tau_v}{\Delta t}\frac{\tau_x}{\Delta t}\frac{T}{4}\left(\frac{d}{ds}\ln\frac{T}{\lambda}\right)^2}.
\end{align}
Since $\dot{\Sigma}$ vanishes in the vanishing limit of the relaxation times, the entropy production $\Sigma$ also vanishes.

\section{finite-time adiabatic process}
\label{finite-time adiabatic process}
To construct the Carnot cycle, we consider the adiabatic process connecting the end of the isothermal process with temperature $T_i$ and the beginning of the other isothermal process with temperature $T_f$.
We introduce the finite-time adiabatic process where the heat flux in Eq.~(\ref{def of the heat flux}) vanishes at any time.
In such a process, we should control the temperature of the heat bath so as to satisfy
\begin{equation}
\label{adiabatic condition}
T(t)=m\sigma_v(t).
\end{equation}
Then, using Eqs.~(\ref{equation of sigma_v}) and (\ref{adiabatic condition}), we find that $\lambda(t)$ should satisfy
\begin{equation}
\label{adiabatic condition for lambda}
    \lambda(t)=-\frac{\dot{T}(t)}{2\sigma_{xv}(t)}.
\end{equation}
Since the heat $Q$ in Eq.~(\ref{def of the heat}) also vanishes in this process, we derive the relation between the output work and the internal energy change in this process as
\begin{equation}
\label{W and Delta E in adiabatic process}
	W=-\Delta E.
\end{equation}
Moreover, since $\dot{Q}$ vanishes, the entropy change of the particle in this process satisfies
\begin{equation}
\label{Delta S and Sigma in adiabatic process}
	\Delta S = \Sigma \geq 0,
\end{equation}
where we used Eq.~(\ref{def of the entropy production}).

Next, we consider how to choose the protocol in the finite-time adiabatic process.
We cannot choose $\Delta t$, $T$, and $\lambda$ independently since they have to satisfy the restriction in Eq.~(\ref{adiabatic condition for lambda}).
We consider the case that the relaxation times are much smaller than $\Delta t$, because the entropy production of the total system vanishes in the vanishing limit of the relaxation times, as mentioned below Eq.~(\ref{entropy production in the small relaxation-times regime}).
Thus, we specify how to choose the protocol when we give a finite $\Delta t$.

In the finite-time adiabatic process with $\Delta t$ given, $\sigma_x$, $\sigma_v$, $\sigma_{xv}$, $T$, and $\lambda$ should satisfy Eqs.~(\ref{equation of sigma_x})--(\ref{equation of sigma_xv}), and (\ref{adiabatic condition}).
In Appendix~\ref{Derivation of the protocol in the adiabatic process}, we derive the protocol of the finite-time adiabatic process from Eqs.~(\ref{equation of sigma_x})--(\ref{equation of sigma_xv}), and (\ref{adiabatic condition}) when we give the time evolution of $\sigma_x(t)$.
Then, $T(t)$ and $\lambda(t)$ are expressed by $\sigma_x(t)$.
To determine $\sigma_x(t)$, we have to give the initial and final values and how to connect them.
We assume that we can arbitrarily connect these values as long as $\sigma_x(t)$ is smooth.
Keeping the small relaxation-times regime of interest in mind, we impose the following condition on the initial and final values of $\sigma_x(t)$:
Since the finite-time adiabatic process continuously connects the isothermal processes, we assume that $T_i$ and $\lambda_i$ ($T_f$ and $\lambda_f$) in the finite-time adiabatic process are the same as those at the end (beginning) of the isothermal process with $T_i$ ($T_f$).
In the small relaxation-times regime, the isothermal process should satisfy Eq.~(\ref{approximation of variables}) at any time.
Thus, the initial and final values of $\sigma_x(t)$ should satisfy 
\begin{equation}
\label{initial and final conditon for sigma_x in the adiabatic process}
    \sigma_{xi}\simeq \frac{T_i}{\lambda_i},\quad \sigma_{xf}\simeq \frac{T_f}{\lambda_f}.
\end{equation}
Below, we only consider the finite-time adiabatic process satisfying the condition in Eq.~(\ref{initial and final conditon for sigma_x in the adiabatic process}).
$\sigma_{xi}$ and $\sigma_{xf}$ are changeable only through $\lambda_i$ and $\lambda_f$ since $T_i$ and $T_f$ are assumed to be given in the Carnot cycle.
Moreover, when we construct the Carnot cycle in the small relaxation-times regime in Sec.~\ref{Carnot cycle in the small relaxation-times regime}, we assume that we can determine the initial and final values of $\lambda(t)$ only in the hot isothermal process among the four processes, which specifically correspond to $\lambda_1$ and $\lambda_2$ in Fig.~\ref{fig:schematic_illustration_2021}, respectively.
Thus, we can give $\lambda_i$ ($\lambda_f$) in the finite-time adiabatic process connecting the end of the hot (cold) isothermal process and the beginning of the cold (hot) isothermal process, where $\lambda_i=\lambda_2$ ($\lambda_f=\lambda_1$) in Fig.~\ref{fig:schematic_illustration_2021}.
This means that we can give only one of $\sigma_{xi}$ and $\sigma_{xf}$ arbitrarily in the finite-time adiabatic process.
The other is determined by the condition for $\Delta t$ as shown below.

The previous study~\cite{PhysRevE.101.032129} considered the finite-time adiabatic process in the overdamped regime of the present model and revealed how $\Delta t$ depends on the time evolution of the protocol and the state of the particle.
We here develop a similar discussion to Ref.~\cite{PhysRevE.101.032129} and reveal the restriction among $\Delta t$ and the five variables ($T(t)$ and $\lambda(t)$ as the protocol and $\sigma_x(t)$, $\sigma_v(t)$, and $\sigma_{xv}(t)$ representing the state of the Brownian particle) in our underdamped system.
From Eqs.~(\ref{Phi}), (\ref{equation of sigma_x})--(\ref{equation of sigma_xv}), and (\ref{adiabatic condition}), we obtain
\begin{align}
    \frac{d\Phi}{dt}=&\frac{d}{dt}(\sigma_x\sigma_v-\sigma_{xv}{}^2)\nonumber\\
    =&2\sigma_{xv}\frac{T}{m}+\sigma_x\left(\frac{2 \gamma T}{m^2} - \frac{2 \gamma}{m}\frac{T}{m} - \frac{2 \lambda}{m}\sigma_{xv}\right)\nonumber\\
    &-2\sigma_{xv}\left(\frac{T}{m} - \frac{\lambda}{m}\sigma_{x} - \frac{ \gamma}{m}\sigma_{xv}\right)\nonumber\\
    =&\frac{2\gamma}{m}\sigma_{xv}{}^2
    =\frac{\gamma}{2m}\left(\frac{d\sigma_x}{dt}\right)^2.
\end{align}
We derive $\Delta \Phi$ in the finite-time adiabatic process as
\begin{align}
	\label{Delta Phi}
	\Delta \Phi
	=&\int^{t_f}_{t_i} dt\ \frac{d\Phi}{dt}\nonumber\\
	=&\frac{\gamma}{2m}\int^{t_f}_{t_i} dt\ \left(\frac{d\sigma_x}{dt}\right)^2\nonumber\\
	=&\frac{\gamma}{2m}\frac{1}{\Delta t}\int^{1}_{0} ds\ \left(\frac{d\sigma_x}{ds}\right)^2,
\end{align}
using $s$ in Eq.~(\ref{normalized time}).
Thus, we obtain $\Delta t$ as
\begin{equation}
\label{Delta t in an adiabatic process}
    \Delta t=\frac{\gamma}{2m}\frac{1}{\Delta \Phi}\int^{1}_{0} ds\ \left(\frac{d\sigma_x}{ds}\right)^2.
\end{equation}
When $\Phi$ at the beginning and end of the finite-time adiabatic process satisfies Eq.~(\ref{approximation of Phi}), Eq.~(\ref{Delta t in an adiabatic process}) can be rewritten as
\begin{equation}
\label{Delta t in an adiabatic process with initial condition}
    \Delta t\simeq\frac{\gamma}{2}\frac{1}{T_f^2/\lambda_f-T_i^2/\lambda_i}\int^{1}_{0} ds\ \left(\frac{d\sigma_x}{ds}\right)^2.
\end{equation}

Giving the finite $\Delta t$ and $\sigma_{x}(t)$ including an undetermined value $\sigma_{xf}$ ($\sigma_{xi}$), we show how to determine $\lambda_f$ ($\lambda_i$) to satisfy Eq.~(\ref{Delta t in an adiabatic process with initial condition}).
Since we can give only one of $\lambda_i$ and $\lambda_f$, we consider each case.
We first consider the case of giving $\lambda_i$.
In the small relaxation-times regime, we obtain $\dot{\sigma}_x(t)=2\sigma_{xv}(t)=O(\tau_{x}(t))$ from Eqs.~(\ref{equation of sigma_x}) and (\ref{sigma xv in the small relaxation times regime}).
Moreover, since $\lambda_i/\lambda(t)$ is finite for any $t$ from the assumption above Eq.~(\ref{ratio of the relaxation times of x}), we obtain $O(\tau_{x}(t))=O(\tau_{xi})$.
Thus, the order of $d\sigma_x(s)/ds=\Delta t\dot{\sigma}_{x}(t)$ is $O(\tau_{xi})$ for any $t$, and the order of the integral in Eq.~(\ref{Delta t in an adiabatic process with initial condition}) is $O(\tau_{xi}^2)$.
To make Eq.~(\ref{Delta t in an adiabatic process with initial condition}) consistent with finite $\Delta t$ in the vanishing limit of $\tau_{xi}$, we impose the condition for $\lambda_i$ and $\lambda_f$ as
\begin{equation}
\label{condition for the adiabatic process_i}
    \frac{T^2_f}{\lambda_f}-\frac{T^2_i}{\lambda_i}=\frac{\alpha}{\gamma}\tau_{xi}^2=\frac{\alpha \gamma}{\lambda_i^2},
\end{equation}
where $\alpha$ is a finite constant to be determined.
Then, from  Eq.~(\ref{Delta t in an adiabatic process with initial condition}), we obtain
\begin{equation}
\label{time with the adiabatic condtion}
    \Delta t\simeq\frac{\lambda_i^2}{2\alpha}\int^{1}_{0} ds\ \left(\frac{d\sigma_x}{ds}\right)^2.
\end{equation}
Because $\lambda_i=\gamma/\tau_{xi}$ and the order of the integral in Eq.~(\ref{time with the adiabatic condtion}) is $O(\tau_{xi}^2)$, the product of $\lambda_i^2$ and the integral remains finite even when we make $\tau_{xi}$ small.
When we give a finite $\Delta t$, $\alpha$ in Eq.~(\ref{time with the adiabatic condtion}) should be positively finite.
Noting that $\sigma_x(t)$ connecting $\sigma_{xi}$ and $\sigma_{xf}$ is given, where $\sigma_{xi}$ is determined by the given $\lambda_i$ from Eq.~(\ref{initial and final conditon for sigma_x in the adiabatic process}) and $\sigma_{xf}$ is to be determined, the result of the integral in Eq.~(\ref{time with the adiabatic condtion}) becomes the function of $\sigma_{xf}$, which can be rewritten in terms of $\lambda_f$, using Eq.~(\ref{initial and final conditon for sigma_x in the adiabatic process}).
Solving Eqs.~(\ref{condition for the adiabatic process_i}) and (\ref{time with the adiabatic condtion}), we can formally obtain $\alpha$ and $\lambda_f$.

We can see that the entropy production of the total system is given by
\begin{align}
\label{the approximation of the entropy production in the adiabatic process}
    \Sigma=&\Delta S
    \simeq\frac{1}{2}\ln\left(\frac{T_f^2}{\lambda_f}\frac{\lambda_i}{T_i^2}\right)
    \simeq\frac{1}{2}\ln\left(1+\frac{\alpha \tau_{xi}}{T_i^2}\right),
\end{align}
from Eqs.~(\ref{def of the entropy}), (\ref{approximation of Phi}), (\ref{Delta S and Sigma in adiabatic process}), and (\ref{condition for the adiabatic process_i}).
Therefore, the entropy production vanishes in the vanishing limit of the relaxation times ($\tau_{xi}\to 0$).

Next, we consider the case of giving $\lambda_f$.
Then, we impose the condition instead of Eq.~(\ref{condition for the adiabatic process_i}) as
\begin{equation}
\label{condition for the adiabatic process_f}
    \frac{T^2_f}{\lambda_f}-\frac{T^2_i}{\lambda_i}=\frac{\alpha'}{\gamma}\tau_{xf}^2=\frac{\alpha' \gamma}{\lambda_f^2},
\end{equation}
where $\alpha'$ is a finite constant to be determined.
By similar discussion to the case of giving $\lambda_i$, we find that $\alpha'$ is positively finite, and obtain
\begin{align}
    \label{time with the adiabatic condtion with tau f}
    \Delta t\simeq& \frac{\lambda_f^2}{2\alpha'}\int^{1}_{0} ds\ \left(\frac{d\sigma_x}{ds}\right)^2,\\
\label{the approximation of the entropy production in the adiabatic process with tau f}
    \Sigma\simeq&-\frac{1}{2}\ln\left(1-\frac{\alpha' \tau_{xf}}{T_f^2}\right).
\end{align}
When we give $\Delta t$, we can obtain $\alpha'$ and $\lambda_{i}$ by solving Eqs.~(\ref{condition for the adiabatic process_f}) and (\ref{time with the adiabatic condtion with tau f}).
From Eq.~(\ref{the approximation of the entropy production in the adiabatic process with tau f}), we can see that the entropy production vanishes in the vanishing limit of the relaxation times ($\tau_{xf}\to 0$).

\section{Carnot cycle in the small relaxation-times regime} 
\label{Carnot cycle in the small relaxation-times regime}
\begin{figure}[t]
  \includegraphics[width=8cm]{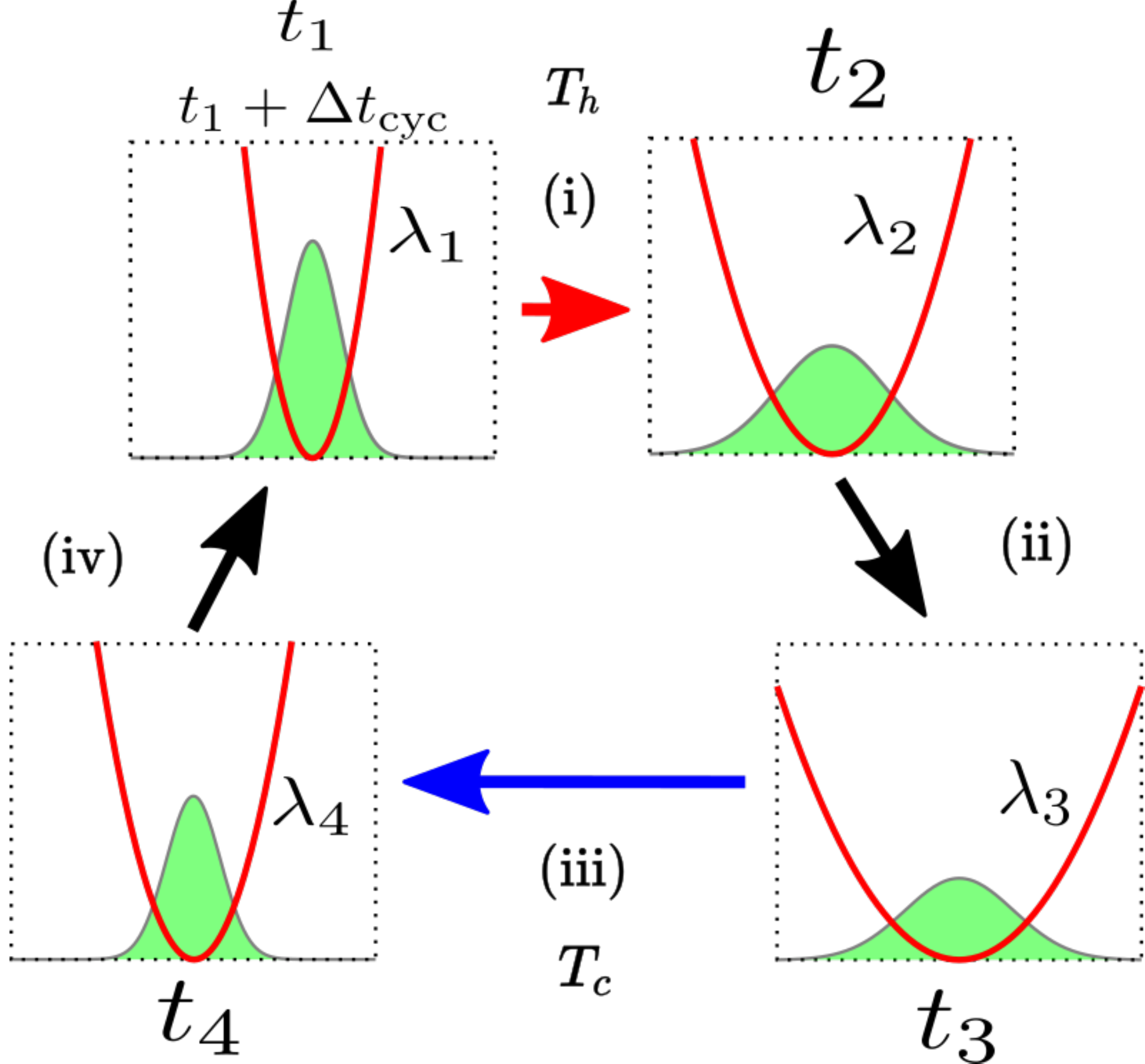}
  \caption{Schematic illustration of the Brownian Carnot cycle.
    In each box, the bottom horizontal line denotes the position coordinate $x$, and the boundary curve of the green filled area denotes the probability distribution of $x$. The red solid line corresponds to the harmonic potential with $\lambda_i$ ($i=1,2,3,4$).
    This cycle is composed of
    (i) the hot isothermal process,
    (ii) the finite-time adiabatic process,
    (iii) the cold isothermal process, and
    (iv) the finite-time adiabatic process.
}
  \label{fig:schematic_illustration_2021}
\end{figure}

We construct a Carnot cycle in the small relaxation-times regime by connecting the hot and cold isothermal processes with the temperature $T_h$ and $T_c$ $(< T_h)$
by the finite-time adiabatic processes (Fig.~\ref{fig:schematic_illustration_2021}).
We assume that $T(t)$ and $\lambda(t)$ are smooth during each process and continuous in the whole cycle including all the switchings between the processes.
Note that $m\sigma_v(t) = T(t)$ holds in the finite-time adiabatic processes from Eq.~(\ref{adiabatic condition}), but that equality may not hold in the isothermal processes.
This may be inconsistent with the continuity assumption of $T(t)$ at the switchings between the isothermal and finite-time adiabatic processes, which means that the finite-time adiabatic processes may not be realized under the continuous $T(t)$.
In the small relaxation-times regime, however, since the approximate equality $\sigma_v \simeq T/m$ in Eq.~(\ref{approximation of variables}) is satisfied in the isothermal processes, we can regard Eq.~(\ref{adiabatic condition}) as being approximately satisfied at the beginning and end of the finite-time adiabatic processes. 
That is, the consistency with the continuity assumption of $T(t)$ recovers in the small relaxation-times regime.
Thus, since we are interested in whether the compatibility of the Carnot efficiency and finite power is possible in the vanishing limit of the relaxation times, we consider only the Carnot cycle in the small relaxation-times regime.

We use the following protocol: (i) The hot isothermal process lasts for $t_1\leq t \leq t_2$.
The temperature of the heat bath satisfies $T(t)=T_h$, and the stiffness $\lambda(t)$ changes from $\lambda_1$ to $\lambda_2$.
(ii) The finite-time adiabatic process connecting the end of the hot isothermal process and the beginning of the cold one lasts for $t_2\leq t \leq t_3$.
The temperature of the heat bath changes from $T_h$ to $T_c$, and the stiffness changes from $\lambda_2$ to $\lambda_3$. 
(iii) The cold isothermal process lasts for $t_3\leq t \leq t_4$.
The temperature of the heat bath satisfies $T(t)=T_c$, and the stiffness $\lambda(t)$ changes from $\lambda_3$ to $\lambda_4$.
(iv) The finite-time adiabatic process connecting the end of the cold isothermal process and the beginning of the hot one lasts for $t_4\leq t \leq t_1+\Delta t_{\rm cyc}$, where $\Delta t_{\rm cyc}$ is the cycle time.
The temperature of the heat bath changes from $T_c$ to $T_h$, and the stiffness changes from $\lambda_4$ to $\lambda_1$.

In the above protocol, we assume that we can choose $\lambda_1$ and $\lambda_2$ arbitrarily and also assume that we choose the time taken for each process to be finite.
When $\lambda_2$ is given, the initial value of $\tau_x$ of the finite-time adiabatic process (ii) is determined from Eq.~(\ref{relaxation time of x}).
Then, since $\Delta t$ in this process is assumed to be finite, the condition of the finite-time adiabatic process in Eq.~(\ref{condition for the adiabatic process_i}) should be satisfied because of the discussion in Sec.~\ref{finite-time adiabatic process}.
Applying Eq.~(\ref{condition for the adiabatic process_i}) to the finite-time adiabatic process (ii), we impose the condition as
\begin{align}
\label{adiabatic condition of h to c}
    &\frac{T_c^2}{\lambda_3}-\frac{T_h^2}{\lambda_2}=\alpha_{h\to c}\frac{\gamma}{\lambda_2^2},
\end{align}
where $\alpha_{h\to c}$ is a finite positive constant.
Here, the indexes ``$h \to c$" and ``$c\to h$" denote the quantities in the finite-time adiabatic processes (ii) and (iv), respectively.
When we give $\sigma_x(t)$ and $\Delta t$ in this process, we obtain the equation for $\alpha_{h\to c}$ and $\lambda_3$ from Eq.~(\ref{time with the adiabatic condtion}).
By solving Eqs.~(\ref{time with the adiabatic condtion}) and (\ref{adiabatic condition of h to c}) simultaneously, we can obtain $\alpha_{h\to c}$ and $\lambda_3$.
In the finite-time adiabatic process (iv), the final value of $\tau_x$ is given since $\lambda_1$ is determined.
Then, since $\Delta t$ in this process is assumed to be finite, the condition in Eq.~(\ref{condition for the adiabatic process_f}) should be satisfied.
Applying Eq.~(\ref{condition for the adiabatic process_f}) to the finite-time adiabatic process (iv), we impose the condition as
\begin{align}
\label{adiabatic condition of c to h}
    &\frac{T_h^2}{\lambda_1}-\frac{T_c^2}{\lambda_4}=\alpha_{c\to h}'\frac{\gamma}{\lambda_1^2},
\end{align}
where $\alpha_{c\to h}'$ is a finite positive constant.
When we give $\sigma_x(t)$ and $\Delta t$ in this process, we can obtain $\lambda_4$ and $\alpha_{c\to h}'$ by solving Eqs.~(\ref{time with the adiabatic condtion with tau f}) and (\ref{adiabatic condition of c to h}) simultaneously.
For convenience, we introduce the function $\phi(t)$ to describe the time evolution of the temperature $T(t)$ as
\begin{equation}
	T(t)=\frac{T_hT_c}{T_h+\left(T_c-T_h\right)\phi(t)}.
\end{equation}
In our Carnot cycle, since we assume that $T(t)$ is continuous, $\phi(t)$ should also be continuous, satisfying
\begin{equation}
\label{time evolution of phi}
\left\{
	\begin{array}{ll}
		\phi(t)=1&(t_1\leq t\leq t_2)\\
		0\leq \phi(t) \leq 1\quad&(t_2\leq t\leq t_3)\\
		\phi(t)=0&(t_3\leq t\leq t_4)\\
		0\leq \phi(t) \leq 1&(t_4\leq t\leq t_1+\Delta t_{\rm cyc}).
	\end{array}
\right.
\end{equation}

\subsection{Construction of the Carnot cycle}

In the hot isothermal process~(i), the time taken for this process is given by
\begin{equation}
\label{Delta t iso h}
\Delta t_{h} \equiv t_2-t_1.
\end{equation}
We choose $\Delta t_{h}$ as a finite value.
When the entropy of the particle changes from $S_1$ to $S_2$ in this process, the entropy change $\Delta S_h$ in this process is given by
\begin{equation}
\label{Delta S h}
	\Delta S_h\equiv S_2-S_1.
\end{equation}
Substituting $T(t)=T_h$ into Eq.~(\ref{def of the entropy production}), the heat $Q_h$ flowing from the heat bath to the particle in this process is expressed by the entropy change of the particle $\Delta S_h$ and entropy production of the total system $\Sigma_h$ in this process as
\begin{equation}
\label{Q h}
	Q_h=T_h\Delta S_h-T_h\Sigma_h.
\end{equation}
Since the entropy production $\Sigma_h$ is nonnegative as seen from Eq.~(\ref{def of the entropy production rate}), we derive the inequality
\begin{equation}
\label{inequality for Q h}
    T_h\Delta S_h\geq Q_h.
\end{equation}
$Q_h>0$ should be satisfied since the heat should flow from the heat bath to the particle in the hot isothermal process in the Carnot cycle useful as a heat engine.
Therefore, $T_h\Delta S_h$ as the upper bound of $Q_h$ in Eq.~(\ref{inequality for Q h}) has to be positive, and we obtain a necessary condition for the hot isothermal process:
\begin{equation}
\label{condition of the hot isothermal process}
	S_2>S_1.
\end{equation}
Since we consider the small relaxation-times regime, we can derive a condition for $\lambda_1$ and $\lambda_2$ from Eq.~(\ref{condition of the hot isothermal process}).
From Eqs.~(\ref{def of the entropy}) and (\ref{approximation of Phi}), the entropy change in the hot isothermal process in this regime can be approximated by
\begin{equation}
\label{the entropy change in the hot isothermal process}
    \Delta S_h\simeq \frac{1}{2}\ln\left(\frac{\lambda_1}{\lambda_2}\right).
\end{equation}
Because of $\Delta S_h>0$, $\lambda_1$ and $\lambda_2$ should satisfy 
\begin{equation}
\label{stiffness 1 and 2}
    \lambda_1>\lambda_2.
\end{equation}
Moreover, $\lambda_1/\lambda_2$ is finite because of the assumption above Eq.~(\ref{ratio of the relaxation times of x}).
Then, because of Eqs.~(\ref{the entropy change in the hot isothermal process}) and (\ref{stiffness 1 and 2}), $\Delta S_h$ in the small relaxation-times regime is positively finite.

In the finite-time adiabatic process~(ii), the entropy change of the particle $\Delta S_{h\to c}$ is equal to the entropy production of the total system $\Sigma_{h\to c}$ because of Eq.~(\ref{Delta S and Sigma in adiabatic process}).
When the entropy of the particle changes from $S_2$ to $S_3$, we obtain
\begin{equation}
\label{Delta S h c}
	\Delta S_{h\to c} \equiv S_3-S_2=\Sigma_{h\to c}.
\end{equation}
From the nonnegatively of $\Sigma_{h\to c}$ in Eq.~(\ref{def of the entropy production rate}), the relation
\begin{equation}
\label{condition of the adiabatic process h c}
	S_3 \geq S_2
\end{equation}
should be satisfied.
The time taken for this process is given by
\begin{equation}
\label{Delta t ad hc}
	\Delta t_{h\to c} \equiv t_3-t_2.
\end{equation}
We choose $\Delta t_{h\to c}$ as a finite value.

In the isothermal process~(iii), we can repeat the discussion similar to the isothermal process (i).
In this process, the entropy of the particle changes from $S_3$ to $S_4$.
Then, the time $\Delta t_{c}$ taken for this process, the entropy change $\Delta S_c$, and the heat $Q_c$ flowing from the heat bath to the particle in this process are given by
\begin{align}
\label{Delta t iso c}
\Delta t_{c} \equiv& t_4-t_3,\\
\label{Delta S c}
	\Delta S_c\equiv& S_4-S_3,\\
\label{Q c}
	Q_c=&T_c\Delta S_c-T_c\Sigma_c,
\end{align}
where $\Sigma_c$ is the entropy production of the total system in this process.
We choose $\Delta t_{c}$ as a finite value.
Note that we cannot determine the sign of $\Delta S_c$ at this point by considering the sign of $Q_c$ unlike the case of the hot isothermal process (i).
We will determine the sign of $\Delta S_c$ later by considering the sum of the entropy change of the particle during one cycle.

\begin{figure}[t]
  \includegraphics[width=6.5cm]{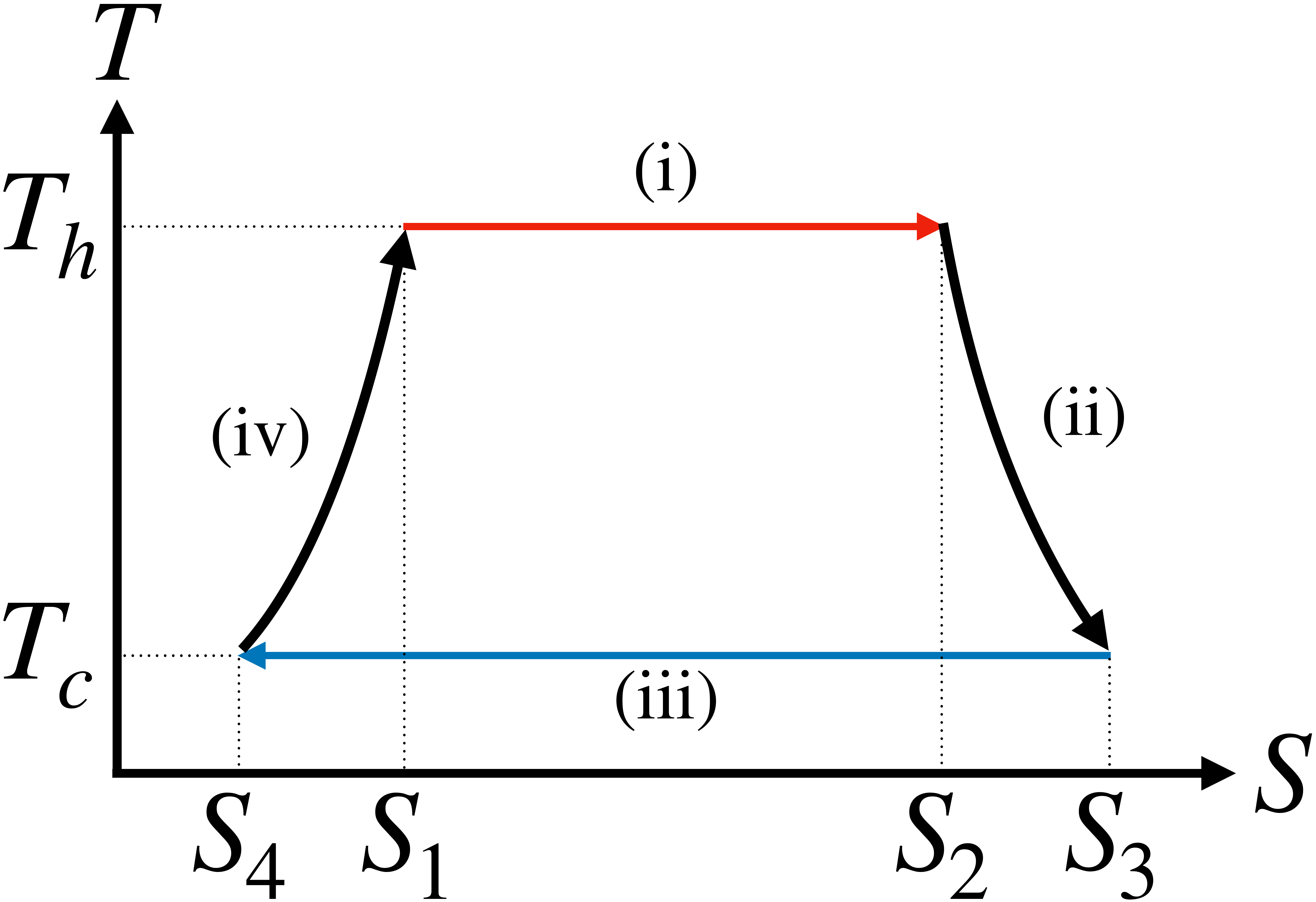}
  \caption{
      The entropy of the Brownian particle and temperature of the heat bath in our Carnot cycle with the finite-time adiabatic processes.
      From the restriction of the finite-time adiabatic processes, the entropy at the ends of the isothermal processes should satisfy $S_2\leq S_3$ and $S_4\leq S_1$.}
  \label{fig:Carnot cycle}
\end{figure}
In the finite-time adiabatic process~(iv), the entropy change of the particle $\Delta S_{c\to h}$ is equal to the entropy production of the total system $\Sigma_{c\to h}$ because of Eq.~(\ref{Delta S and Sigma in adiabatic process}).
For the cycle to close, the entropy of the particle should change from $S_4$ to $S_1$, which leads to
\begin{equation}
\label{Delta S c h}
	\Delta S_{c\to h} \equiv S_1-S_4=\Sigma_{c\to h}.
\end{equation}
Since the entropy production $\Sigma_{c\to h}$ is nonnegative, $S_4$ and $S_1$ should satisfy
\begin{equation}
\label{condition of the adiabatic process c h}
	S_1 \geq S_4.
\end{equation}
The time taken for this process is given by
\begin{equation}
\label{Delta t ad ch}
	\Delta t_{c\to h} \equiv (t_1+\Delta t_{\rm cyc})-t_4.
\end{equation}
We choose $\Delta t_{c\to h}$ as  a finite value.

After one cycle, the system returns to the initial state.
Then, we obtain the trapezoidlike $T$-$S$ diagram of our cycle in Fig.~\ref{fig:Carnot cycle} from Eqs.~(\ref{condition of the hot isothermal process}), (\ref{condition of the adiabatic process h c}), and (\ref{condition of the adiabatic process c h}).
The sum of the entropy change of the particle satisfies
\begin{align}
\label{entropy change per cycle}
	&\Delta S_h+\Delta S_{h\to c}+\Delta S_c+\Delta S_{c\to h}=0.
\end{align}
Using Eqs.~(\ref{Delta S h c}), (\ref{Delta S c h}), and (\ref{entropy change per cycle}), we obtain
\begin{equation}
\label{entropy change per cycle 2}
	\Delta S_c=-(\Delta S_h+\Sigma_{h\to c}+\Sigma_{c\to h}).
\end{equation}
Using Eqs.~(\ref{def of the entropy}) and (\ref{approximation of Phi}), we find that the entropy change of the particle in the cold isothermal process in the small relaxation-times regime can be approximated by
\begin{equation}
\label{the entropy change in the cold isothermal process}
    \Delta S_c\simeq \frac{1}{2}\ln\left(\frac{\lambda_3}{\lambda_4}\right)<0,
\end{equation}
where the last inequality comes from $\Delta S_h+\Sigma_{h\to c}+\Sigma_{c\to h}> 0$ and Eq.~(\ref{entropy change per cycle 2}).
Then, we obtain
\begin{equation}
\label{stiffness 3 and 4}
    \lambda_3<\lambda_4.
\end{equation}
From the first law of the thermodynamics, we derive the output work per cycle as
\begin{align}
\label{work per cycle}
	W =& Q_h+Q_c\nonumber\\
	=& T_h\Delta S_h + T_c\Delta S_c -T_h\Sigma_h-T_c\Sigma_c\\
	=&(T_h-T_c)\Delta S_h -T_h\Sigma_h-T_c(\Sigma_{h\to c}+\Sigma_c+\Sigma_{c\to h})\nonumber,
\end{align}
using Eqs.~(\ref{Q h}), (\ref{Q c}), and (\ref{entropy change per cycle 2}).
Since the entropy production of the total system in each process is nonnegative, the work $W$ has the upper bound $W_0$ as
\begin{equation}
\label{inequality of the work}
	W_0\equiv (T_h-T_c)\Delta S_h\geq W.
\end{equation}
Since $T_h-T_c$ and $\Delta S_h$ are finite, $W_0$ is also finite.
When the entropy production in each process vanishes, we obtain $W=W_0$ from Eq.~(\ref{work per cycle}).
By using Eqs.~(\ref{Delta t iso h}), (\ref{Delta t ad hc}), (\ref{Delta t iso c}), and (\ref{Delta t ad ch}), the time taken for each process satisfies
\begin{equation}
\label{cycle time}
	\Delta t_{\rm cyc}=\Delta t_h+\Delta t_{h\to c}+\Delta t_c+\Delta t_{c\to h}.
\end{equation}
Since the time taken for each process is finite, $\Delta t_{\rm cyc}$ is finite.
Using Eqs.~(\ref{Carnot efficiency}), (\ref{inequality for Q h}), and (\ref{inequality of the work}), we obtain the conditions for the efficiency $\eta$ and power $P$ of our cycle as
\begin{align}
\label{def of the efficiency}
	\eta\equiv& \frac{W}{Q_h}
	\leq\frac{W_0}{T_h\Delta S_h}
	=\frac{(T_h-T_c)\Delta S_h}{T_h\Delta S_h} 
	=\eta_{\rm C},
\end{align}
\begin{align}
\label{def of the power}
	P\equiv& \frac{W}{\Delta t_{\rm cyc}}
	\leq \frac{W_0}{\Delta t_{\rm cyc}}
	=\frac{(T_h-T_c)\Delta S_h}{\Delta t_{\rm cyc}}
	\equiv P_0,
\end{align}
where $P_0$ is the power when $W=W_0$ is satisfied.
When the entropy productions $\Sigma_h$, $\Sigma_c$, $\Sigma_{h\to c}$, and $\Sigma_{c\to h}$ vanish, $Q_h$ and $W$ approach $T_h\Delta S_h$ in Eq.~(\ref{inequality for Q h}) and $W_0$ in Eq.~(\ref{inequality of the work}), respectively.
Then, the efficiency approaches the Carnot efficiency $\eta_{\rm C}$ and the power approaches $P_0$.

\section{theoretical analysis}
\label{theoretical analysis}
\subsection{Trade-off relation}
\label{Trade-off relation}
We show the trade-off relation between the efficiency $\eta$ and power $P$ in our cycle.
From Eq.~(\ref{def of the entropy production rate}), we obtain the following inequality:
\begin{align}	
	\dot{\Sigma}\geq& \frac{\dot{Q}^2}{\gamma T \sigma_v},
\end{align}
or equivalently,
\begin{equation}
\label{inequality of the heat flux}
	|\dot{Q}|\leq \sqrt{\gamma T \sigma_v\dot{\Sigma}}.
\end{equation}
Since the finite-time adiabatic processes satisfy $\dot{Q}=0$, $Q_h$ can be written as
\begin{equation}
	Q_h=\int^{t_2}_{t_1}dt\ \dot{Q}=\int^{t_1+\Delta t_{\rm cyc}}_{t_1}dt\ \phi(t)\dot{Q}(t),
\end{equation}
where we used Eq.~(\ref{time evolution of phi}).
Then, by using the Cauchy-Schwarz inequality, we derive the inequality for $Q_h$ as
\begin{align}
\label{inequality of Q h}
	Q_h^2=&\left(\int^{t_1+\Delta t_{\rm cyc}}_{t_1}dt\ \phi\dot{Q}\right)^2\nonumber\\
	\leq&\left(\int^{t_1+\Delta t_{\rm cyc}}_{t_1}dt\ \phi|\dot{Q}|\right)^2\nonumber\\
	\leq&\left(\int^{t_1+\Delta t_{\rm cyc}}_{t_1}dt\ \phi\sqrt{\gamma T \sigma_v\dot{\Sigma}}\right)^2\nonumber\\
	\leq&\Delta t_{\rm cyc}T_c^2\chi\Sigma_{\rm cyc},
\end{align}
using Eq.~(\ref{inequality of the heat flux}), where $\Sigma_{\rm cyc}$ is the entropy production of the total system per cycle and $\chi$ is defined as
\begin{equation}
\label{def of the chi}
\chi \equiv \frac{\gamma}{T_c^2\Delta t_{\rm cyc}}\int^{t_1+\Delta t_{\rm cyc}}_{t_1}dt\ \phi(t)^2 T(t) \sigma_v(t).
\end{equation}
Multiplying both sides of Eq.~(\ref{inequality of Q h}) by $W/(\Delta t_{\rm cyc}Q_h^2)$, we derive the inequality for the power in Eq.~(\ref{def of the power}) as
\begin{align}
\label{inequality for the power}
	P\leq \frac{T_c^2\eta}{Q_h}\chi\Sigma_{\rm cyc},
\end{align}
using Eq.~(\ref{def of the efficiency}).
Since the entropy change of the particle vanishes after one cycle, the entropy production of the total system per cycle is equal to that of the heat bath.
Thus, the entropy production per cycle relates to the efficiency as
\begin{align}
\label{Sigma with efficiency}
	\Sigma_{\rm cyc}
	=&-\frac{Q_h}{T_h}-\frac{Q_c}{T_c}\nonumber\\
	=&\frac{Q_h}{T_c}(\eta_{\rm C}-\eta),
\end{align}
using Eqs.~(\ref{Carnot efficiency}), (\ref{work per cycle}), and (\ref{def of the efficiency}).
From Eq.~(\ref{Sigma with efficiency}), it turns out that the efficiency becomes the Carnot efficiency when $\Sigma_{\rm cyc}$ vanishes.
By using Eqs.~(\ref{inequality for the power}) and (\ref{Sigma with efficiency}), we can derive the trade-off relation between the efficiency and power:
\begin{align}
\label{trade-off relation}
	P\leq \chi T_c \eta(\eta_{\rm C}-\eta).
\end{align}
This inequality is the same as Eq.~(111) in Ref.~\cite{PhysRevE.103.042125}, where we applied the method based on Ref.~\cite{PhysRevE.97.062101}.

\subsection{Compatibility of the Carnot efficiency and finite power}
We show the compatibility of the Carnot efficiency and finite power in the vanishing limit of the relaxation times in our cycle.
If the entropy productions of the total system in all the processes vanish, the entropy production per cycle vanishes.
Then, we can achieve the Carnot efficiency because of Eq.~(\ref{Sigma with efficiency}), and the power also approaches a finite $P_0$ in Eq.~(\ref{def of the power}).
Thus, we evaluate the entropy production in each process in the small relaxation-times regime.

From Eq.~(\ref{entropy production in the small relaxation-times regime}), the entropy productions in the isothermal processes are given by
\begin{align}
\label{entropy production h in the small relaxation-times regime}
	\Sigma_h\simeq&\frac{1}{T_h}\int^{1}_{0}ds_h \frac{\frac{\tau_{v}}{\Delta t_h}\left(\frac{dQ}{ds_h}\right)^2
    + \frac{\tau_{x}}{\Delta t_h}\frac{T_h^2}{4}\left(\frac{d}{ds_h}\ln\frac{T_h}{\lambda}\right)^2}
  {T_h-\frac{\tau_{x}}{\Delta t_h}\frac{\tau_{v}}{\Delta t_h}\frac{T_h}{4}\left(\frac{d}{ds_h}\ln\frac{T_h}{\lambda}\right)^2},\\
  \label{entropy production c in the small relaxation-times regime}
  \Sigma_c\simeq&\frac{1}{T_c}\int^{1}_{0}ds_c \frac{\frac{\tau_{v}}{\Delta t_c}\left(\frac{dQ}{ds_c}\right)^2
    + \frac{\tau_{x}}{\Delta t_c}\frac{T_c^2}{4}\left(\frac{d}{ds_c}\ln\frac{T_c}{\lambda}\right)^2}
  {T_{c}-\frac{\tau_{x}}{\Delta t_c}\frac{\tau_{v}}{\Delta t_c}\frac{T_c}{4}\left(\frac{d}{ds_c}\ln\frac{T_c}{\lambda}\right)^2},
\end{align}
where $s_{h}$ and $s_{c}$ are the normalized times in the corresponding isothermal processes:
\begin{equation}
    \label{normalized time for isothermal processes}
    s_{h}\equiv\frac{t-t_1}{\Delta t_{h}},\quad s_{c}\equiv\frac{t-t_3}{\Delta t_{c}}.
\end{equation}
In the isothermal processes, $\lambda$ is not constant because of Eqs.~(\ref{stiffness 1 and 2}) and (\ref{stiffness 3 and 4}), while $T$ is constant. 
Then, from Eq.~(\ref{heat flux with small relaxation times}), we find that $dQ/ds_{h}$ in Eq.~(\ref{entropy production h in the small relaxation-times regime}) and $dQ/ds_{c}$ in Eq.~(\ref{entropy production c in the small relaxation-times regime}) are finite when $\lambda$ changes smoothly except when $\lambda$ takes extremal values.
Thus, in the vanishing limit of the relaxation times, the integrand in Eqs.~(\ref{entropy production h in the small relaxation-times regime}) and (\ref{entropy production c in the small relaxation-times regime}) vanishes, and $\Sigma_h$ and $\Sigma_c$ also vanish.

Since we choose $\Delta t_{h\to c}$ and $\Delta t_{c\to h}$ as finite values, $\alpha_{h\to c}$ and $\alpha'_{c\to h}$ in Eqs.~(\ref{adiabatic condition of h to c}) and (\ref{adiabatic condition of c to h}) are positively finite because of the discussion below Eq.~(\ref{time with the adiabatic condtion}).
By applying Eqs.~(\ref{the approximation of the entropy production in the adiabatic process}) and (\ref{the approximation of the entropy production in the adiabatic process with tau f}) to the present finite-time adiabatic processes (ii) and (iv), respectively, we derive the entropy productions of the total system in these processes as
\begin{align}
\label{entropy production h to c in the small relaxation-times regime}
    \Sigma_{h\to c}\simeq& \frac{1}{2}\ln\left(1+\frac{\alpha_{h\to c} \tau_{x}(t_2)}{T_h^2}\right),\\
    \label{entropy production c to h in the small relaxation-times regime}
    \Sigma_{c\to h}\simeq& -\frac{1}{2}\ln\left(1-\frac{\alpha'_{c\to h} \tau_{x}(t_1)}{T_h^2}\right).
\end{align}
These entropy productions vanish in the vanishing limit of the relaxation times.

Note that there may exist an entropy production at the switchings between the isothermal and finite-time adiabatic processes.
As shown in Appendix~\ref{entropy production at the switching}, this entropy production is caused by the discontinuity of the time derivative of $T(t)$ and $\lambda(t)$ at the switchings, although we assume that $T(t)$ and $\lambda(t)$ are continuous.
However, we can also show that this entropy production is negligible in the small relaxation-times regime.

Using Eqs.~(\ref{entropy production h in the small relaxation-times regime}), (\ref{entropy production c in the small relaxation-times regime}), (\ref{entropy production h to c in the small relaxation-times regime}), and (\ref{entropy production c to h in the small relaxation-times regime}), we obtain the entropy production of the total system per cycle:
\begin{align}
\label{entropy production per cycle}
    \Sigma_{\rm cyc}=&\Sigma_h+\Sigma_c+\Sigma_{h\to c}+\Sigma_{c\to h}\nonumber\\
    \simeq&\frac{1}{T_h}\int^{1}_{0}ds_h \frac{\frac{\tau_{v}}{\Delta t_h}\left(\frac{dQ}{ds_h}\right)^2
    + \frac{\tau_{x}}{\Delta t_h}\frac{T_h^2}{4}\left(\frac{d}{ds_h}\ln\frac{T_h}{\lambda}\right)^2}
  {T_h-\frac{\tau_{x}}{\Delta t_h}\frac{\tau_{v}}{\Delta t_h}\frac{T_h}{4}\left(\frac{d}{ds_h}\ln\frac{T_h}{\lambda}\right)^2}\nonumber\\
  &+\frac{1}{T_c}\int^{1}_{0}ds_c \frac{\frac{\tau_{v}}{\Delta t_c}\left(\frac{dQ}{ds_c}\right)^2
    + \frac{\tau_{x}}{\Delta t_c}\frac{T_c^2}{4}\left(\frac{d}{ds_c}\ln\frac{T_c}{\lambda}\right)^2}
  {T_{c}-\frac{\tau_{x}}{\Delta t_c}\frac{\tau_{v}}{\Delta t_c}\frac{T_c}{4}\left(\frac{d}{ds_c}\ln\frac{T_c}{\lambda}\right)^2}\nonumber\\
  &+\frac{1}{2}\ln\left(1+\frac{\alpha_{h\to c} \tau_{x}(t_2)}{T_h^2}\right)\nonumber\\
  &-\frac{1}{2}\ln\left(1-\frac{\alpha'_{c\to h} \tau_{x}(t_1)}{T_h^2}\right).
\end{align}
From the discussion below Eqs.~(\ref{normalized time for isothermal processes}) and (\ref{entropy production c to h in the small relaxation-times regime}), the entropy productions in all the processes vanish in the vanishing limit of the relaxation times, and $\Sigma_{\rm cyc}$ also vanishes in this limit.
Then, the heat $Q_h$ in Eq.~(\ref{Q h}) and work $W$ in Eq.~(\ref{work per cycle}) become $T_h\Delta S_h$ in Eq.~(\ref{inequality for Q h}) and $W_0$ in Eq.~(\ref{inequality of the work}), respectively.
Thus, the efficiency in Eq.~(\ref{def of the efficiency}) approaches the Carnot efficiency, and the power in Eq.~(\ref{def of the power}) approaches $P_0$.
Since $\Delta t_{\rm cyc}$ in Eq.~(\ref{cycle time}) and $W_0$ are finite, $P_0$ is finite.
Although this may seem to be inconsistent with the trade-off relation in Eq.~(\ref{trade-off relation}), we below show that there is no inconsistency.

In the small relaxation-times regime, since $\sigma_v$ is approximated by Eq.~(\ref{approximation of variables}), $\chi$ in Eq.~(\ref{def of the chi}) is approximated by
\begin{align}
    \label{approximation of chi}
    \chi\simeq&\frac{1}{\tau_v\Delta t_{\rm cyc}T_c^2}\int^{t_1+\Delta t_{\rm cyc}}_{t_1}dt\ \phi^2 T^2\nonumber\\
    =&\frac{1}{\tau_v}\left(\frac{1}{T_c^2}\int^{1}_{0}ds_{\rm cyc}\ \phi^2 T^2\right)\nonumber\\
    =&\frac{C}{\tau_v},
\end{align}
where $s_{\rm cyc}\equiv (t-t_1)/\Delta t_{\rm cyc}$, and $C$ is a constant defined as
\begin{equation}
    C\equiv\frac{1}{T_c^2}\int^{1}_{0}ds_{\rm cyc}\ \phi^2 T^2.
\end{equation}
Since $T$ is finite and $\phi$ satisfies Eq.~(\ref{time evolution of phi}), $C$ is finite.
From Eq.~(\ref{approximation of chi}), $\chi$ turns out to diverge in the vanishing limit of the relaxation times.

Next, we consider the right-hand side of Eq.~(\ref{inequality for the power}).
In the small relaxation-times regime, $\chi\Sigma_{\rm cyc}$ in Eq.~(\ref{inequality for the power}) is approximated by
\begin{align}
\label{chi Sigma}
\chi\Sigma_{\rm cyc}\simeq&\frac{C}{T_h}\int^{1}_{0}ds_h \frac{\frac{1}{\Delta t_h}\left(\frac{dQ}{ds_h}\right)^2
    + \frac{\tau_{x}}{\tau_v\Delta t_h}\frac{T_h^2}{4}\left(\frac{d}{ds_h}\ln\frac{T_h}{\lambda}\right)^2}
  {T_h-\frac{\tau_{x}}{\Delta t_h}\frac{\tau_{v}}{\Delta t_h}\frac{T_h}{4}\left(\frac{d}{ds_h}\ln\frac{T_h}{\lambda}\right)^2}\nonumber\\
  &+\frac{C}{T_c}\int^{1}_{0}ds_c \frac{\frac{1}{\Delta t_c}\left(\frac{dQ}{ds_c}\right)^2
    + \frac{\tau_{x}}{\tau_v\Delta t_c}\frac{T_c^2}{4}\left(\frac{d}{ds_c}\ln\frac{T_c}{\lambda}\right)^2}
  {T_c-\frac{\tau_{x}}{\Delta t_c}\frac{\tau_{v}}{\Delta t_c}\frac{T_c}{4}\left(\frac{d}{ds_c}\ln\frac{T_c}{\lambda}\right)^2}\nonumber\\
  &+\frac{C}{2\tau_v}\ln\left(1+\frac{\alpha_{h\to c} \tau_{x}(t_2)}{T_h^2}\right)\nonumber\\
  &-\frac{C}{2\tau_v}\ln\left(1-\frac{\alpha'_{c\to h} \tau_{x}(t_1)}{T_h^2}\right)
\end{align}
from Eqs.~(\ref{entropy production per cycle}) and (\ref{approximation of chi}).
As shown below Eq.~(\ref{normalized time for isothermal processes}), $dQ/ds_{h}$ and $dQ/ds_{c}$ in the isothermal processes are finite except when $\lambda$ takes extremal values.
Then, the first term of the numerator of the integrand in the first and second terms of Eq.~(\ref{chi Sigma}) is positively finite even in the vanishing limit of the relaxation times since $\Delta t_{h}$ and $\Delta t_{c}$ are finite.
Thus, the first and second terms in Eq.~(\ref{chi Sigma}) do not vanish in the vanishing limit of the relaxation times, which means that $\chi\Sigma_{\rm cyc}$ does not vanish in this limit, while $\Sigma_{\rm cyc}$ vanishes and the efficiency approaches the Carnot efficiency because of Eq.~(\ref{Sigma with efficiency}).
Therefore, the right-hand side of the trade-off relation in Eq.~(\ref{inequality for the power}) does not vanish, which means that the finite power is allowed.
In fact, the power also approaches the finite $P_0$ in this limit.
Thus, the compatibility of the Carnot efficiency and finite power is achievable by taking the vanishing limit of the relaxation times in our Brownian Carnot cycle with arbitrary temperature difference.

\section{Numerical simulation}
\label{numerical simulation}
We show numerical results of the compatibility of the Carnot efficiency and finite power in the vanishing limit of the relaxation times.
In this simulation, we solved Eqs.~(\ref{equation of sigma_x})--(\ref{equation of sigma_xv}) numerically by the fourth-order Runge-Kutta method.
Our specific protocol in the isothermal processes is given by
\begin{align}
  \label{protocol in the overdamped regime}
    T(t)=&\ T_h \hspace{3mm} (t_1\leq t\leq t_2),\nonumber\\
    T(t)=&\ T_c \hspace{3mm} (t_3\leq t\leq t_4),\\
    \lambda_h(t)=&\ \frac{T_h}{w_1\left[1+\left(\sqrt{w_2/w_1}-1\right)\frac{t-t_1}{\Delta t_h}\right]^2}
    \hspace{3mm} (t_1\leq t\leq t_2),\nonumber\\
    \lambda_c(t)=&\ \frac{T_c}{w_3\left[1+\left(\sqrt{w_4/w_3}-1\right)\frac{t-t_3}{\Delta t_c}\right]^2}
    \hspace{2mm}(t_3\leq  t\leq  t_4)\nonumber,
\end{align}
where $\lambda_{h}(t)$ and $\lambda_{c}(t)$ are the time evolution of  $\lambda(t)$ in the isothermal processes with temperatures $T_{h}$ and $T_{c}$, respectively, and $w_{j}$ ($j=1,2,3,4$) are positive parameters.
This protocol is inspired by the optimal protocol in the overdamped Brownian Carnot cycle with the instantaneous adiabatic processes~\cite{Schmiedl_2007,PhysRevE.96.062107} and used in our previous study~\cite{PhysRevE.103.042125}.
From Eqs.~(\ref{relaxation time of x}) and (\ref{protocol in the overdamped regime}), we obtain
\begin{equation}
\label{w with tau x}
    w_j=\frac{T(t_j)}{\lambda_j} = \frac{T(t_j)}{\gamma}\tau_x(t_j).
\end{equation}
Note that, using Eqs.~(\ref{approximation of variables}) and (\ref{w with tau x}), we obtain
\begin{equation}
\label{sigma_x at the edges}
    \sigma_{x}(t_j)\simeq w_j
\end{equation}
in the small relaxation-times regime.
From Eqs.~(\ref{stiffness 1 and 2}) and (\ref{w with tau x}), $w_2/w_1>1$ should be satisfied.
For all simulations, we fixed $w_2/w_1=2.0$, since we can choose $w_1$ and $w_2$ arbitrarily, corresponding to the assumption that we can choose $\lambda_1$ and $\lambda_2$ arbitrarily, as mentioned above Eq.~(\ref{adiabatic condition of h to c}). 
Note that $w_2/w_1$ should be finite since $\lambda_1/\lambda_2$ is finite as shown below Eq.~(\ref{stiffness 1 and 2}).
We also fixed $T_{c} = 1.0$, $\Delta t_{h}=\Delta t_{c}=1.0$, and $\gamma = 1.0$ and varied $w_1$, $m$, and the temperature difference $T_h-T_c$ (or equivalently, the temperature $T_{h}$).
Note that Eqs.~(\ref{w with tau x}) and (\ref{sigma_x at the edges}) satisfy the condition in Eq.~(\ref{initial and final conditon for sigma_x in the adiabatic process}).

We have to consider the finite-time adiabatic processes to determine $w_3$ and $w_4$.
By using Eqs.~(\ref{adiabatic condition of h to c}), (\ref{adiabatic condition of c to h}), and (\ref{w with tau x}), we obtain
\begin{align}
    \label{w_3}
    w_3=\frac{T_h}{T_c}w_2+\frac{\gamma\alpha_{h\to c}}{T_h^2T_c}w_2^2,\\
    \label{w_4}
    w_4=\frac{T_h}{T_c}w_1-\frac{\gamma\alpha'_{c\to h}}{T_h^2T_c}w_1^2,
\end{align}
where $\alpha_{h\to c}$ and $\alpha_{c\to h}'$ are constants.
Below, we explain how to determine $w_3$, $w_4$, $\alpha_{h\to c}$, and $\alpha_{c\to h}'$ from Eqs.~(\ref{time with the adiabatic condtion}), (\ref{time with the adiabatic condtion with tau f}), (\ref{w_3}), and (\ref{w_4}) when $\Delta t_{h\to c}$ and $\Delta t_{c\to h}$ are given.
First, we give $\sigma_{x}(t)$ in the finite-time adiabatic processes to obtain the protocol.
From Eq.~(\ref{sigma_x at the edges}), we give $\sigma_{x,h\to c}(t)$ satisfying $\sigma_{xi,h\to c}=w_2$ and $\sigma_{xf,h\to c}=w_3$.
Although $\sigma_{xi,h\to c}$ is given since we can give $w_2$, $\sigma_{xf,h\to c}$ is undetermined. 
Similarly, we give $\sigma_{x,c\to h}(t)$ satisfying $\sigma_{xi,c\to h}=w_4$ and $\sigma_{xf,c\to h}=w_1$, where $\sigma_{xf,c\to h}$ is given and $\sigma_{xi,c\to h}$ is undetermined.
Moreover, we assume $\Delta t_{h\to c}= \Delta t_{c\to h}= 1.0$.
Using $\sigma_x(t)$, $\Delta t_{h\to c}$, and $\Delta t_{c\to h}$, we can obtain the equations for $w_3$, $w_4$, $\alpha_{h\to c}$, and $\alpha_{c\to h}'$ from Eqs.~(\ref{time with the adiabatic condtion}) and (\ref{time with the adiabatic condtion with tau f}) in our cycle.
Then, solving those equations together with Eqs.~(\ref{w_3}) and (\ref{w_4}), we can determine $w_3$, $w_4$, $\alpha_{h\to c}$, and $\alpha_{c\to h}'$.
Note that we can choose $\Delta t_{h\to c}$ and $\Delta t_{c\to h}$ arbitrarily as long as they are finite, although we set $\Delta t_{h\to c}=\Delta t_{c\to h}= 1.0$ in our simulation for simplicity.

In the finite-time adiabatic processes, we give
\begin{align}
\label{sigma x in the numerical simulation}
    \sigma_x(t)=&\left\{
    \begin{array}{ll}
        w_{h\to c}(t)& (t_2\leq t \leq t_3)\\
        w_{c\to h}(t)& (t_4\leq t\leq t_1+\Delta t_{\rm cyc}),
    \end{array}
    \right.\\
\label{w h to c}
    w_{h\to c}(t)\equiv &\ w_2+(w_3-w_2)(t-t_2)/\Delta t_{h\to c}, \\
\label{w c to h}    
    w_{c\to h}(t)\equiv &\ w_4+(w_1-w_4)(t-t_4)/\Delta t_{c\to h}
\end{align}
to obtain the protocol.
Then, from Eqs.~(\ref{stiffness in adiabatic process}) and (\ref{T in adiabatic process}) in Appendix~\ref{Derivation of the protocol in the adiabatic process}, we find that the protocol is obtained as
\begin{align}
    \label{optimized temperature}
    T_{h\to c}(t)=&\left(1-\frac{w_3\frac{t-t_2}{\Delta t_{h\to c}}}{ w_{h\to c}(t)}\right)T_h+\frac{w_3\frac{t-t_2}{\Delta t_{h\to c}}}{w_{h\to c}(t)}T_c
    \nonumber\\
    &\hspace{3.5cm}(t_2\leq t\leq t_3),\nonumber\\
    T_{c\to h}(t)=&\left(1-\frac{w_{1}\frac{t-t_4}{\Delta t_{c\to h}}}{ w_{c\to h}(t)}\right)T_c +\frac{w_{1}\frac{t-t_4}{\Delta t_{c\to h}}}{ w_{c\to h}(t)}T_h\\
    &\hspace{2.5cm}(t_4\leq t\leq t_1+\Delta t_{\rm cyc}),\nonumber
\end{align} 
\begin{align}
    \label{optimized stiffness}
    \lambda_{h\to c}(t)=&\frac{2T_{h\to c}(t)-\gamma\frac{\left(w_{3}-w_{2}\right)}{\Delta t_{h\to c}}}{2 w_{h\to c}(t)}\quad(t_2\leq t\leq t_3),\nonumber\\
    \lambda_{c\to h}(t)=&\frac{2T_{c\to h}(t)-\gamma\frac{\left(w_{1}-w_{4}\right)}{\Delta t_{c\to h}}}{2 w_{c\to h}(t)}\\
    &\hspace{2.5cm}(t_4\leq t\leq t_1+\Delta t_{\rm cyc}).\nonumber
\end{align}
Note that the approximate equality in Eq.~(\ref{sigma_x at the edges}) becomes equality only in the vanishing limit of the relaxation times.
Although we derived the protocol in the adiabatic processes by regarding $\sigma_x(t_{j})$ as $w_j$ ($j=1,2,3,4$) in Eqs.~(\ref{sigma x in the numerical simulation})--(\ref{w c to h}), the equality in Eq.~(\ref{sigma_x at the edges}) is not satisfied exactly in the simulation.
Thus, the time evolution of $\sigma_x(t)$ realized by solving Eqs.~(\ref{equation of sigma_x})--(\ref{equation of sigma_xv}) with the protocol in Eqs.~(\ref{optimized temperature}) and (\ref{optimized stiffness}) is not exactly the same as the given $\sigma_x(t)$ in Eq.~(\ref{sigma x in the numerical simulation}).
However, their difference becomes small when the relaxation times are sufficiently small.
Moreover, because the domains of $w_{h\to c}$ and $w_{c\to h}$ are $t_2\leq t\leq t_3$ and $t_4\leq t\leq t_1+\Delta t_{\rm cyc}$, respectively, they satisfy
\begin{equation}
\label{inequality for w}
    0\leq \frac{w_3\frac{t-t_2}{\Delta t_{h\to c}}}{w_{h\to c}(t)},\ \frac{w_1\frac{t-t_4}{\Delta t_{c\to h}}}{w_{c\to h}(t)} \leq 1.
\end{equation}
Thus, $T_c\leq T_{h\to c}(t), T_{c\to h}(t)\leq T_h$ is satisfied because of Eq.~(\ref{optimized temperature}).
Then, $T_{h\to c}(t)$ and $T_{c\to h}(t)$ are finite at any time even in the vanishing limit of $w_j$ ($j=1,2,3,4$).
By using Eq.~(\ref{sigma x in the numerical simulation}), we can calculate the integral in Eqs.~(\ref{time with the adiabatic condtion}) and (\ref{time with the adiabatic condtion with tau f}).
Then, $\Delta t_{h\to c}$ and $\Delta t_{c\to h}$ in the small relaxation-times regime satisfy
\begin{align}
\label{Delta t h to c}
    \Delta t_{h\to c}\simeq& \frac{T_h^2}{2\alpha_{h\to c}}\left(\frac{w_3}{w_2}-1\right)^2,\\
\label{Delta t c to h}
    \Delta t_{c\to h}\simeq& \frac{T_h^2}{2\alpha_{c\to h}'}\left(\frac{w_4}{w_1}-1\right)^2.
\end{align}
From Eqs.~(\ref{w_3}) and (\ref{w_4}), we obtain
\begin{align}
    \label{ratio of w2 and w3}
    \frac{w_3}{w_2}=\frac{T_h}{T_c}+\frac{\gamma\alpha_{h\to c}}{T_h^2T_c}w_2\simeq\frac{T_h}{T_c},\\
    \label{ratio of w1 and w4}
    \frac{w_4}{w_1}=\frac{T_h}{T_c}-\frac{\gamma\alpha'_{c\to h}}{T_h^2T_c}w_1\simeq\frac{T_h}{T_c},
\end{align}
where the last approximate equalities hold because $w_j$ in Eq.~(\ref{w with tau x}) is sufficiently small in the small relaxation-times regime and $\alpha_{h\to c}$ and $\alpha'_{c\to h}$ should be finite for any value of the relaxation times.
Then, Eqs.~(\ref{Delta t h to c}) and (\ref{Delta t c to h}) become
\begin{align}
    \label{Delta t h c with temp}
    \Delta t_{h\to c}\simeq& \frac{T_h^2}{2\alpha_{h\to c}}\left(\frac{T_h}{T_c}-1\right)^2,\\
    \label{Delta t c h with temp}
    \Delta t_{c\to h}\simeq& \frac{T_h^2}{2\alpha_{c\to h}'}\left(\frac{T_h}{T_c}-1\right)^2.
\end{align}
Since we choose $\Delta t_{h\to c}=\Delta t_{c\to h}= 1.0$, $\alpha_{h\to c}$ and $\alpha'_{c\to h}$ are given by
\begin{align}
\label{alpha h to c}
    \alpha_{h\to c}\simeq& \frac{T_h^2}{2}\left(\frac{T_h}{T_c}-1\right)^2,\\
\label{alpha c to h}
    \alpha_{c\to h}'\simeq& \frac{T_h^2}{2}\left(\frac{T_h}{T_c}-1\right)^2.
\end{align}
From Eqs.~(\ref{w_3}), (\ref{w_4}), (\ref{alpha h to c}), and (\ref{alpha c to h}), we obtain $w_3$ and $w_4$.

By numerically calculating the integrals in Eqs.~(\ref{def of the work}) and (\ref{def of the heat}) from the solution to Eqs.~(\ref{equation of sigma_x})--(\ref{equation of sigma_xv}), we obtained the heat $Q_h$ and $Q_c$ in Eqs.~(\ref{Q h}) and (\ref{Q c}) and the work $W$ in Eq.~(\ref{work per cycle}).
Using the heat and work, we also calculated the efficiency $\eta=W/Q_h$ in Eq.~(\ref{def of the efficiency}) and power $P=W/\Delta t_{\rm cyc}$ in Eq.~(\ref{def of the power}).
In this simulation, we choose the initial condition as $\sigma_x(t_1)=w_1$, $\sigma_v(t_1)=T_h/m$, and $\sigma_{xv}(t_1)=0$.
Before starting to measure the thermodynamic quantities, we waited until the system settled down to a steady cycle. 
Therefore, our results are insensitive to the initial condition.
When we consider the small relaxation-times regime in the present protocol, we should take the limit of $w_1 \to 0$ and the limit of $m \to 0$ for the following reasons. In the limit of $w_1 \to 0$, we can see that all $w_j$ vanish from $w_2/w_1=2.0$ and Eqs.~(\ref{w_3}) and (\ref{w_4}).
Then, $w_{h\to c}(t)$ in Eq.~(\ref{w h to c}) and $w_{c\to h}(t)$ in Eq.~(\ref{w c to h}) vanish at any time.
Since $T(t)$ is finite at any time in the limit of $w_1 \to 0$ as shown below Eq.~(\ref{inequality for w}), $\lambda(t)$ in Eqs.~(\ref{protocol in the overdamped regime}) and (\ref{optimized stiffness}) diverges and the relaxation time of position $\tau_x(t)$ in Eq.~(\ref{relaxation time of x}) vanishes at any time.
Moreover, in the limit of $m\to 0$, the relaxation time of velocity $\tau_{v}$ in Eq.~(\ref{relaxation time of v}) vanishes.
Note that in the numerical simulations, we selected a time step smaller than the relaxation times. 
Specifically, we set the time step of the Runge-Kutta method as $\min(w_1,m)\times 10^{-2}$ because of $\tau_x(t_1)=\gamma w_1/T_h$ and $\tau_v=m/\gamma$.

We here confirm that the present protocol satisfies the assumption of finite $T(t')/T(t)$ and $\lambda(t')/\lambda(t)$, as mentioned above Eq.~(\ref{ratio of the relaxation times of x}), where $t$ and $t'$ are any times in a process.
Since $T_c\leq T(t)\leq T_h$ is satisfied at any time, $T(t')/T(t)$ is finite in all the processes.
Moreover, from Eqs.~(\ref{ratio of w2 and w3}) and (\ref{ratio of w1 and w4}) and the finite $w_2/w_1$, we find that $w_3/w_1$ and $w_4/w_1$ are also finite.
Then, using Eqs.~(\ref{protocol in the overdamped regime}), (\ref{w h to c}), (\ref{w c to h}), and (\ref{optimized stiffness}), we can confirm that $\lambda(t')/\lambda(t)$ is finite in all the processes even in the vanishing limit of the relaxation times.

\begin{figure}[t]
  \includegraphics[width=6.5cm]{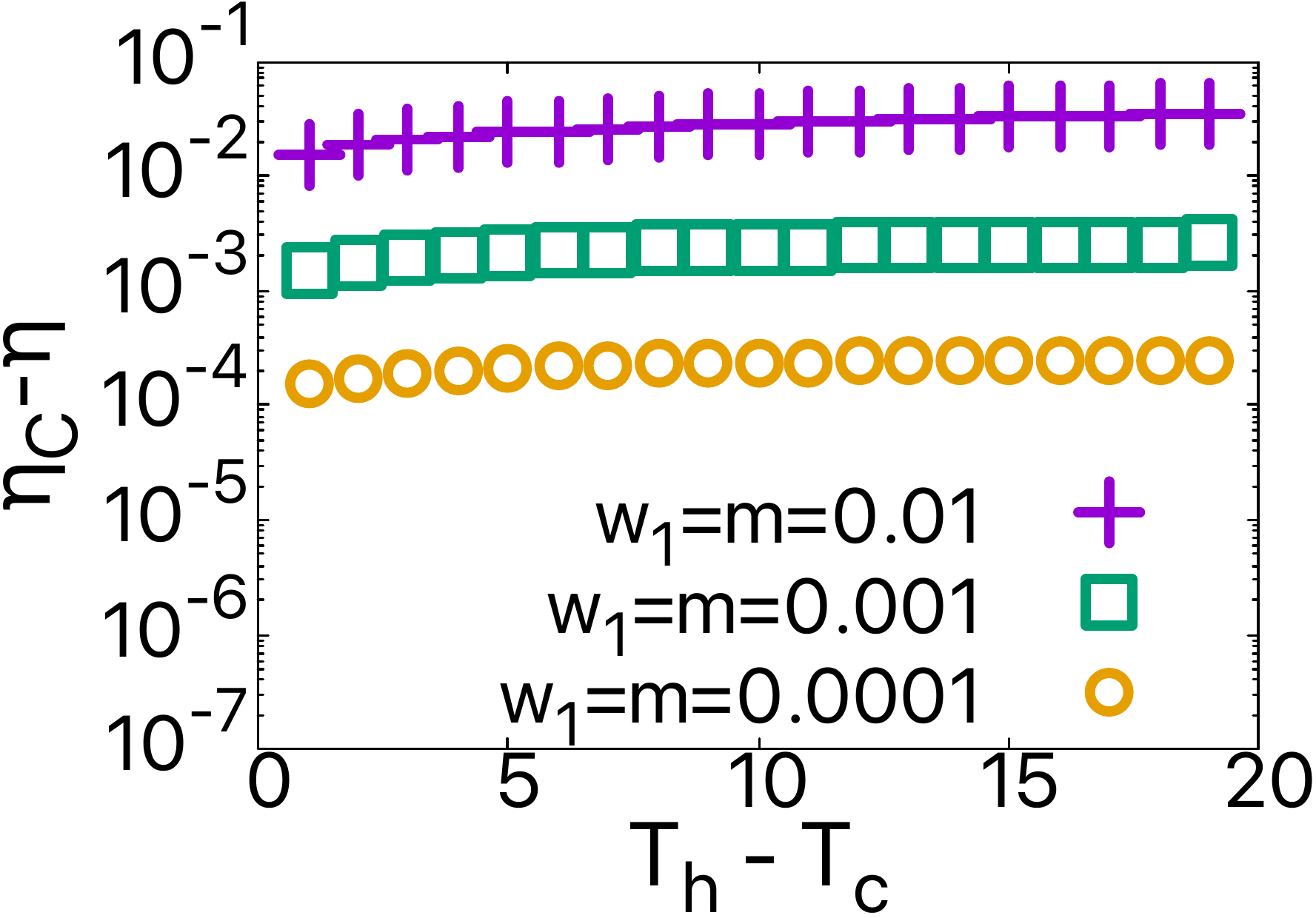}
  \caption{
      The difference between the Carnot efficiency and our efficiency measured with the protocol in Eqs.~(\ref{protocol in the overdamped regime}), (\ref{optimized temperature}), and (\ref{optimized stiffness}), when we vary the relaxation times $\tau_x(t_1)$ and $\tau_v$, which are proportional to $w_1$ and $m$, respectively.
      In these simulations, we set $w_1=m=10^{-2}$ (purple plus), $w_1=m=10^{-3}$ (green square), and $w_1=m=10^{-4}$ (orange circle).
      The efficiency appears to approach the Carnot efficiency in the vanishing limit of $w_1$ (or $\tau_{x}$) and $m$ (or $\tau_{v}$).
  }
  \label{fig:efficiency}
\end{figure}
\begin{figure}[t]
  \includegraphics[width=6.5cm]{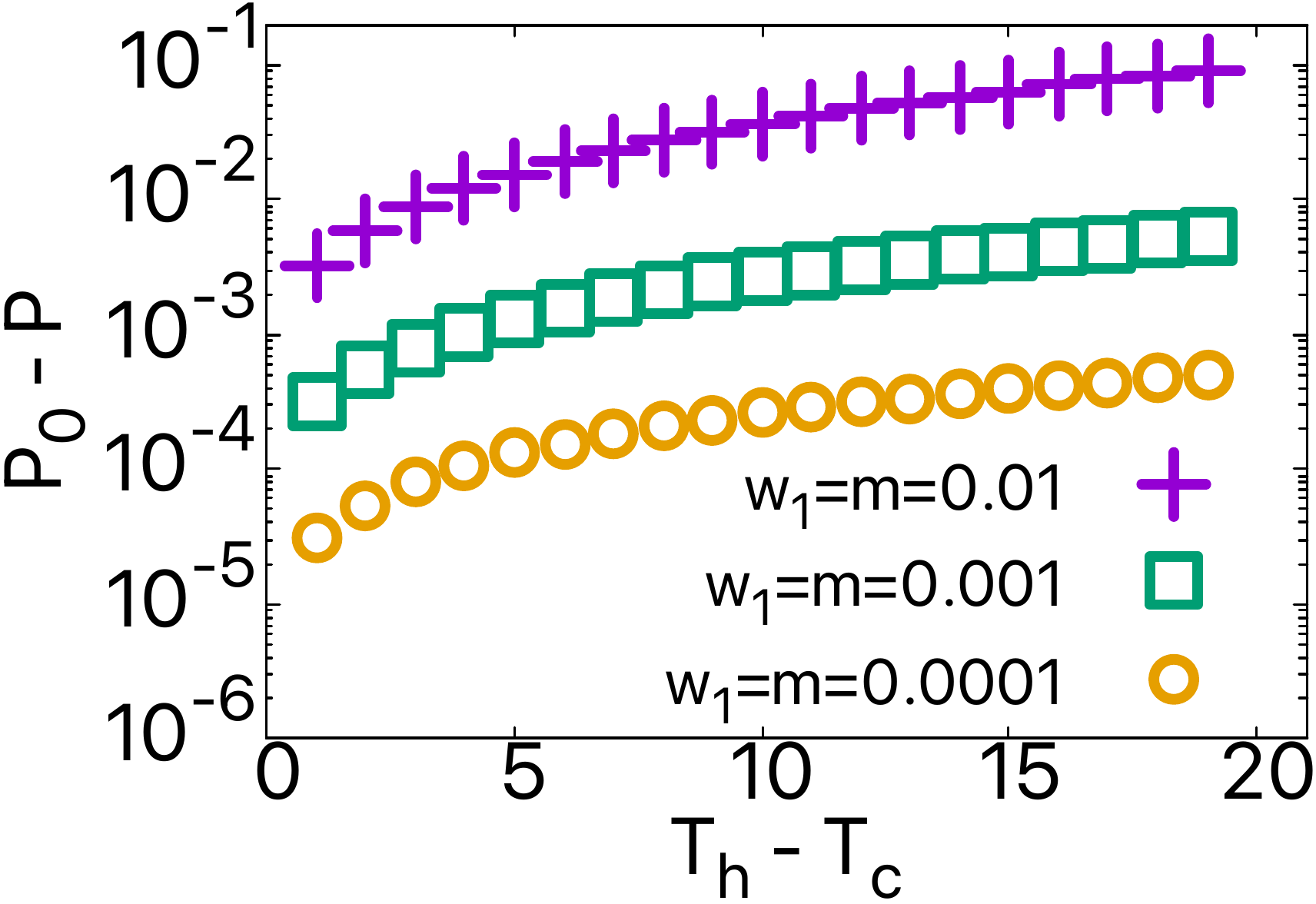}
  \caption{
      The power in Eq.~(\ref{def of the power}) of our cycle, corresponding to Fig.~\ref{fig:efficiency}.
      The power approaches $P_0$ in Eq.~(\ref{P 0 in the numerical simulation}) in the vanishing limit of $w_1$ (or $\tau_{x}$) and $m$ (or $\tau_{v}$) for any temperature difference.
}
  \label{fig:power}
\end{figure}
\begin{figure}[t]
  \includegraphics[width=6.5cm]{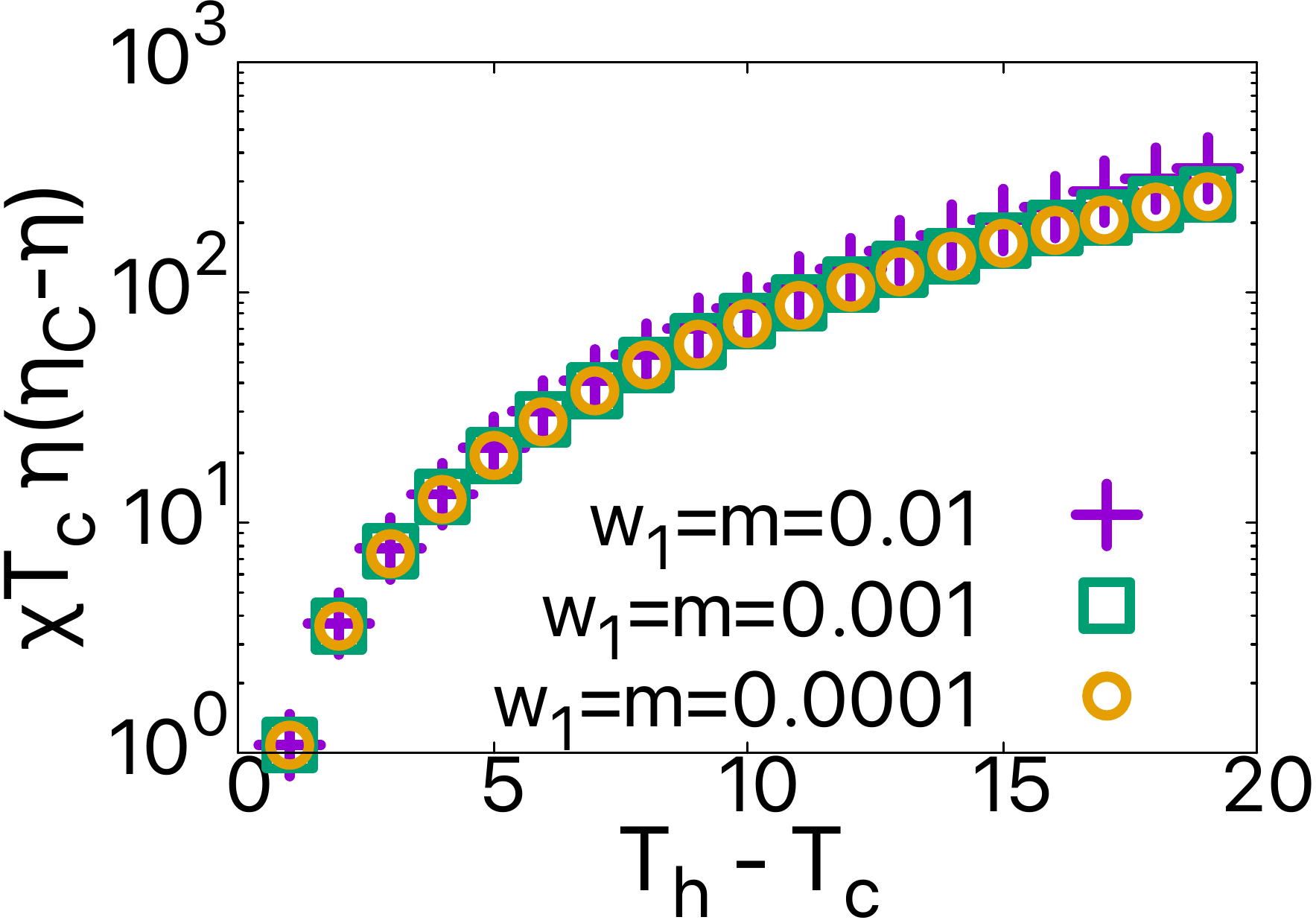}
  \caption{
      The right-hand side of Eq.~(\ref{trade-off relation}) corresponding to Fig.~\ref{fig:efficiency}. 
      Although the efficiency approaches the Carnot efficiency in the limit of $w_1,m\to 0$ ($\tau_{x},\tau_{v} \to 0$) in Fig.~\ref{fig:efficiency}, $\chi T_c\eta(\eta_{\rm C}-\eta)$ remains almost unchanged.
      This means that $\chi$ diverges in this limit.
  }
  \label{fig:bound}
\end{figure}

Figure~\ref{fig:efficiency} shows the difference between the Carnot efficiency and our efficiency measured with the protocol in Eqs.~(\ref{protocol in the overdamped regime}), (\ref{optimized temperature}), and (\ref{optimized stiffness}).
In our previous study~\cite{PhysRevE.103.042125}, the Carnot efficiency is achievable only in the small temperature-difference regime even when we take the vanishing limit of the relaxation times.
In contrast to that, we can see that the efficiency approaches the Carnot efficiency in this limit even when the temperature difference is large.

Figure~\ref{fig:power} shows the difference between $P_0$ in Eq.~(\ref{def of the power}) and $P$ corresponding to Fig.~\ref{fig:efficiency}.
In the small relaxation-times regime, we obtain 
\begin{equation}
\label{Delta S h in the small relaxation times regime}
    \Delta S_h\simeq \frac{1}{2}\ln\left(\frac{w_2}{w_1}\right)=\frac{1}{2}\ln2,
\end{equation}
using $w_2/w_1=2.0$ and Eqs.~(\ref{the entropy change in the hot isothermal process}) and (\ref{w with tau x}).
In the vanishing limit of the relaxation times, since the approximate equality in Eq.~(\ref{approximation of variables}) becomes equality, the approximate equality in Eq.~(\ref{Delta S h in the small relaxation times regime}) also becomes equality.
Then, we obtain $W_0 = (T_h-T_c)\ln2/2$ because of Eq.~(\ref{inequality of the work}).
Since we use $\Delta t_h=\Delta t_c=\Delta t_{h\to c}=\Delta t_{c\to h}=1.0$, we have $\Delta t_{\rm cyc}= 4.0$ in our simulation.
Thus, we obtain
\begin{equation}
\label{P 0 in the numerical simulation}
    P_0=\frac{\ln2}{8}(T_h-T_c),
\end{equation}
from Eq.~(\ref{def of the power}).
From Fig.~\ref{fig:power}, we can see that the power approaches $P_0$ in the vanishing limit of the relaxation times for arbitrary temperature difference.
Thus, the power turns out to be finite.

Figure~\ref{fig:bound} shows the upper bound of the power in Eq.~(\ref{trade-off  relation}).
We can see that the upper bound of the power remains finite even when the efficiency approaches the Carnot efficiency as shown in Fig.~\ref{fig:efficiency}.
This means that $\chi$ in Eq.~(\ref{def of the chi}) diverges in the vanishing limit of the relaxation times, and also means that the finite power is allowed.

\begin{figure}[t]
  \includegraphics[width=6.5cm]{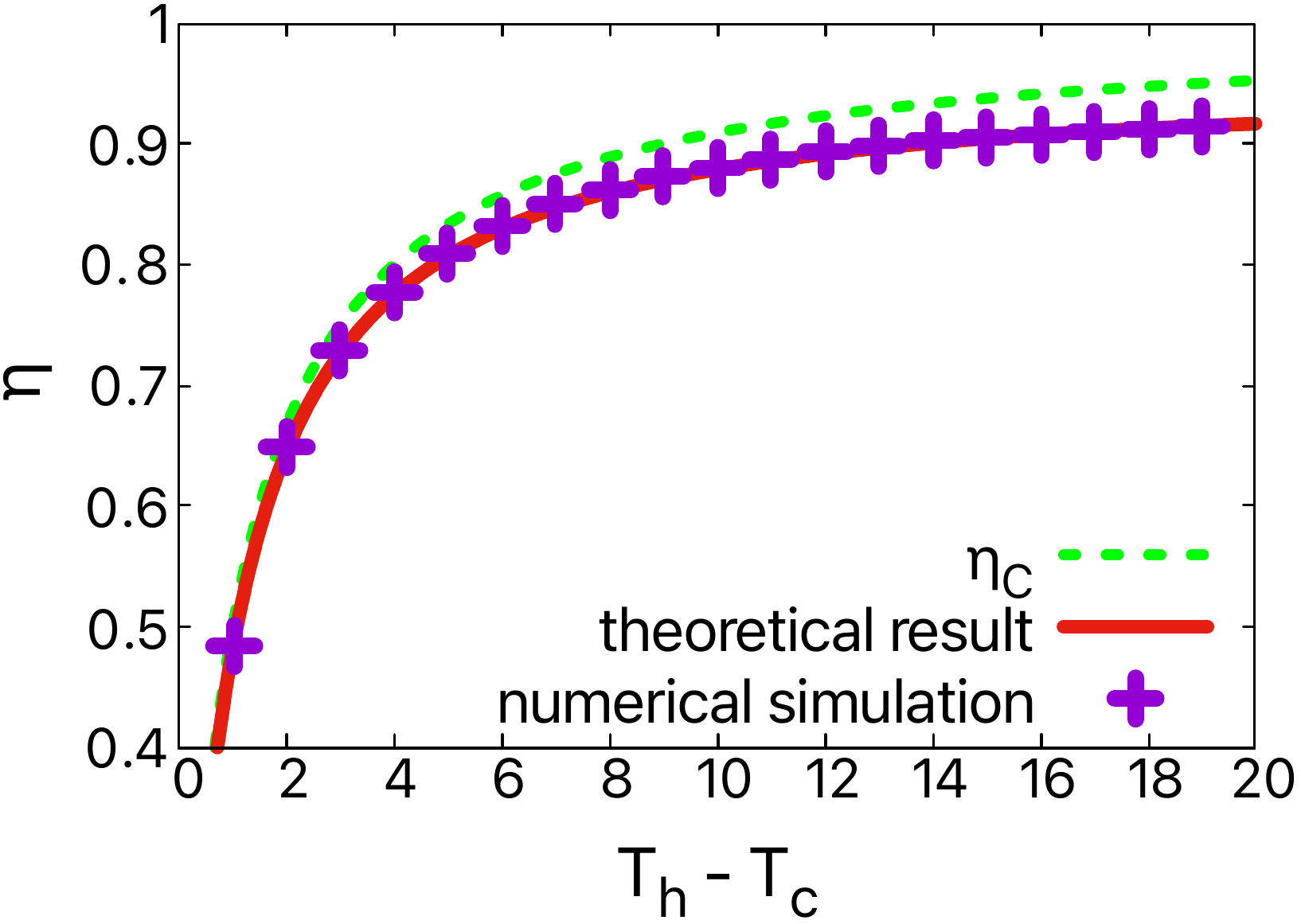}
   \includegraphics[width=6.5cm]{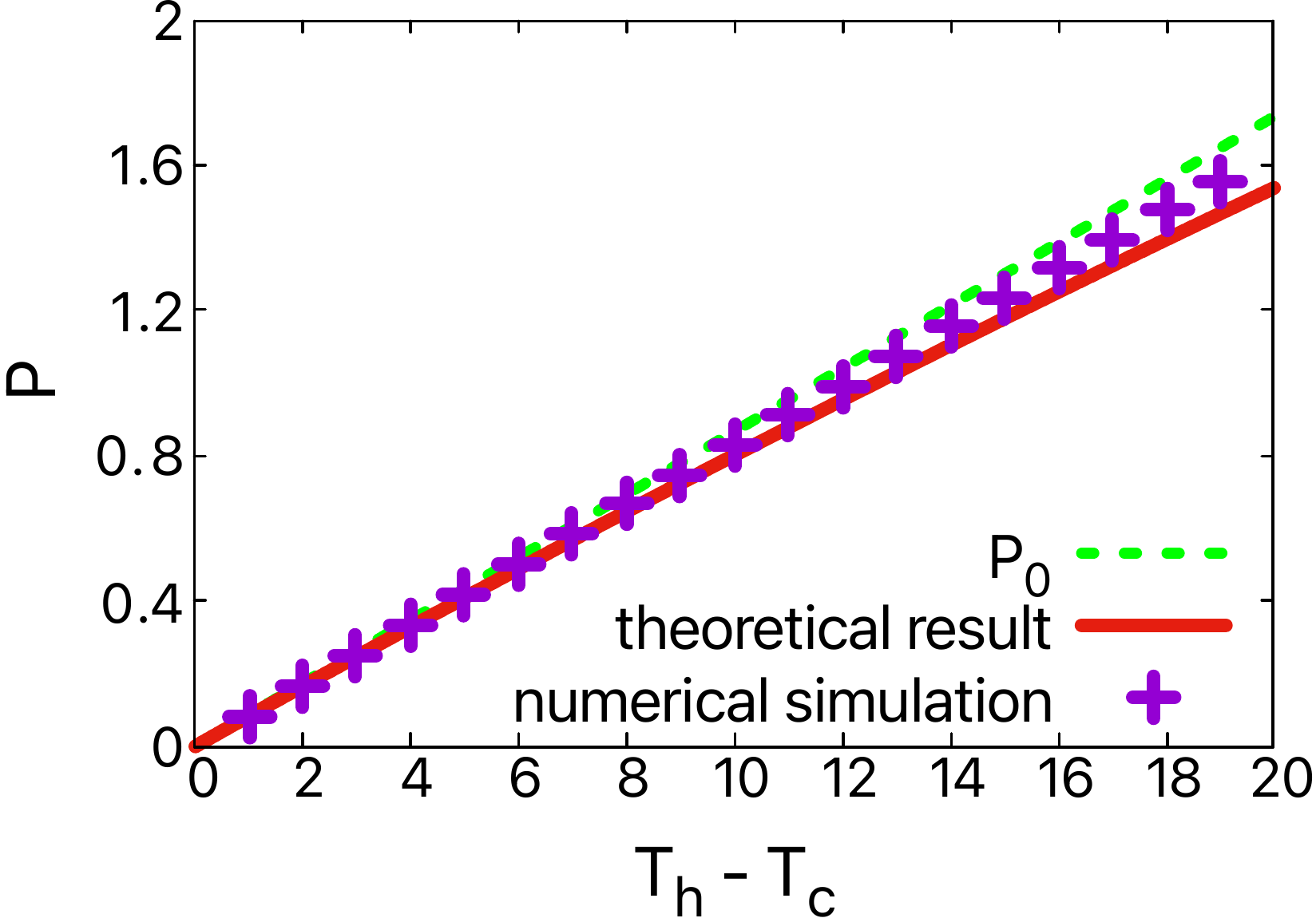}
  \caption{
      Efficiency~(upper figure) and power~(lower figure) derived from the numerical simulations in Figs.~\ref{fig:efficiency} and \ref{fig:power} (purple plus) and theoretical analysis (red solid line).
      We set $w_1=m=10^{-2}$. 
      Although the relaxation times corresponding to these parameters are not very small among the parameters used in Fig.~\ref{fig:efficiency}, the theoretical results and numerical simulations show a good agreement. 
      We have confirmed a better agreement with smaller parameters (data not shown).}
  \label{fig:compare}
\end{figure}

In Fig.~\ref{fig:compare}, we compare the results of the numerical simulation with the theoretical analysis derived in Sec.~\ref{theoretical analysis} for the efficiency and power. 
To obtain the theoretical results in Fig.~\ref{fig:compare}, we calculated the entropy productions in Eqs.~(\ref{entropy production h in the small relaxation-times regime}), (\ref{entropy production c in the small relaxation-times regime}), (\ref{entropy production h to c in the small relaxation-times regime}), and (\ref{entropy production c to h in the small relaxation-times regime}).
Note that we used Eq.~(\ref{heat flux with small relaxation times}) to calculate $\dot{Q}$ in Eqs.~(\ref{entropy production h in the small relaxation-times regime}) and (\ref{entropy production c in the small relaxation-times regime}) from the time derivative of the protocol in Eq.~(\ref{protocol in the overdamped regime}).
Moreover, we set $\Delta S_h=\ln 2/2$ in the theoretical analysis as in the numerical simulation.
Then, we derived the heat in the hot isothermal process and work from Eqs.~(\ref{Q h}) and (\ref{work per cycle}).
Using the heat and work, we can obtain the efficiency in Eq.~(\ref{def of the efficiency}) and power in Eq.~(\ref{def of the power}).
The numerical simulation and theoretical analysis in Fig.~\ref{fig:compare} show good agreement.
Thus, this result shows the validity of the theoretical analysis in Sec.~\ref{theoretical analysis}.
From Figs.~\ref{fig:efficiency}--\ref{fig:bound}, we can see that the compatibility of the Carnot efficiency with finite power is achievable without breaking the trade-off relation.

\section{Summary}
\label{summary and discussion}
We studied the relaxation-times dependence of the efficiency and power in an underdamped Brownian Carnot cycle with the finite-time adiabatic processes~\cite{PhysRevE.101.032129,Plata_2020} and time-dependent harmonic potential.
We showed that the compatibility of the Carnot efficiency and finite power is achievable in the vanishing limit of the relaxation times in our cycle.
In our previous study~\cite{PhysRevE.103.042125}, we showed that the compatibility of the Carnot efficiency and finite power is possible only in the small temperature-difference regime in the Brownian Carnot cycle with the instantaneous adiabatic processes.
In this paper, we considered the finite-time adiabatic processes and represented the entropy production in terms of the relaxation times in the small relaxation-times regime.
Then, the entropy production vanishes in the vanishing limit of the relaxation times.
We constructed the Carnot cycle with the finite-time adiabatic processes in the small relaxation-times regime.
By the theoretical analysis of our cycle, we derived the trade-off relation and showed that in the vanishing limit of the relaxation times, the entropy production per cycle vanishes, in other words, the efficiency approaches the Carnot efficiency.
Then, we also showed that the finite power is achievable without breaking the trade-off relation in Eq.~(\ref{trade-off relation}).
Moreover, we confirmed that our theoretical analysis agrees with the results of our numerical simulation.
We finally note that we can use other protocols satisfying the assumption above Eq.~(\ref{ratio of the relaxation times of x}) and continuity at the switchings between the processes instead of the present protocol in our simulation.

\appendix
\renewcommand{\appendixname}{APPENDIX}
\section{DERIVATION OF EQ.~(\ref{approximation of variables})}
\label{Derivation of Eq ref (approximation of variables)}
We show that the variables $\sigma_{x}$, $\sigma_{v}$, and $\sigma_{xv}$ behave like Eq.~(\ref{approximation of variables}) in the small relaxation-times regime when the time-dependent temperature $T(t)$ and stiffness $\lambda(t)$ satisfy the assumption above Eq.~(\ref{ratio of the relaxation times of x}).
For the above purpose, we first show that the variables $\sigma_x$, $\sigma_v$, and $\sigma_{xv}$ relax toward Eq.~(\ref{approximation of variables}) when the temperature $T$ and stiffness $\lambda$ are constant.
After that, we consider the case that the temperature $T(t)$ and stiffness $\lambda(t)$ depend on time and show that these variables satisfy Eq.~(\ref{approximation of variables}).

We assume that a thermodynamic process lasts for $t_{i} \leq t \leq t_{f}$ and we thus have $\Delta t= t_f-t_i$ in Eq.~(\ref{def of time interval}).
The temperature $T$ and stiffness $\lambda$ are assumed to be constant.
When we set $\sigma_{x}(t_{i})=\sigma_{x0}$, $\sigma_{v}(t_{i})=\sigma_{v0}$, and $\sigma_{xv}(t_{i})=\sigma_{xv0}$ as an initial condition, we can solve Eqs.~(\ref{equation of sigma_x})--(\ref{equation of sigma_xv})
using the Laplace transform~\cite{PhysRevE.97.022131}, and we can obtain $\sigma_{x}$ and $\sigma_{v}$ as follows:
\begin{align}
  \label{sigma_x with constant lambda}
    \sigma_{x}(t) =&\frac{T}{\lambda}
    +\frac{m}{\lambda} D_{1}e^{-\frac{\gamma}{m}(t-t_{i})}\nonumber\\
    &+\frac{\left(\gamma + m\omega^{*}\right)^2}{4\lambda^2}D_{2}e^{-\left(\frac{\gamma}{m} - \omega^{*}\right)(t-t_{i})}\nonumber\\
    & +\frac{\left(\gamma - m\omega^{*}\right)^2}{4\lambda^2}D_{3}e^{-\left(\frac{\gamma}{m} + \omega^{*}\right)(t-t_{i})},\\
  \label{sigma_v with constant lambda}
    \sigma_{v}(t)=&\frac{T}{m}
    +D_{1}e^{-\frac{\gamma}{m} (t-t_{i})}
    +D_{2}e^{-\left(\frac{\gamma}{m} - \omega^{*}\right)(t-t_{i})}\nonumber\\
    &+D_{3}e^{-\left(\frac{\gamma}{m} + \omega^{*}\right)(t-t_{i})},
\end{align}
where
\begin{align}
  \label{omega star}
  \omega^{*} \equiv& \frac{\gamma}{m}\sqrt{1-4\frac{m\lambda}{\gamma^2}},
\end{align}
\begin{align}
  \label{D1}
  D_{1} \equiv &\frac{\lambda}{m{\omega^{*}}^2}
  \left(4\frac{T}{m}-2\sigma_{v0}
  -2\frac{\lambda}{m}\sigma_{x0}
  -2\frac{\gamma}{m}\sigma_{xv0}
  \right),
\end{align}
\begin{align}
  \label{D2}
    D_{2} \equiv &-\frac{1}{2{\omega^{*}}^2}\left[\frac{\gamma T}{m^2}\left(\frac{\gamma}{m}-\omega^{*}\right)
      + \left( 2\frac{\lambda}{m}-\frac{\gamma^2}{m^2}+\frac{\gamma}{m}\omega^{*}\right)\sigma_{v0}\right.\nonumber\\
      &-2\frac{\lambda^2}{m^2}\sigma_{x0}
      \left.+2\frac{\lambda}{m}\left(-\frac{\gamma}{m}+\omega^{*}\right)\sigma_{xv0}\right],
\end{align}
\begin{align}
  \label{D3}
    D_{3} \equiv &-\frac{1}{2{\omega^{*}}^2}\left[\frac{\gamma T}{m^2}
      \left(\frac{\gamma}{m}+\omega^{*}\right)
      + \left(2\frac{\lambda}{m}-\frac{\gamma^2}{m^2}-\frac{\gamma}{m}\omega^{*}\right)\sigma_{v0}\right.\nonumber\\
      &-2\frac{\lambda^2}{m^2}\sigma_{x0}
      \left.+2\frac{\lambda}{m}\left(-\frac{\gamma}{m} - \omega^{*}\right)\sigma_{xv0}\right].
\end{align}
We can also derive $\sigma_{xv}$ using Eqs.~(\ref{equation of sigma_x}) and (\ref{sigma_x with constant lambda}).
Note that since the exponential terms in Eqs.~(\ref{sigma_x with constant lambda}) and (\ref{sigma_v with constant lambda}) vanish as $t \to \infty$, $\sigma_x$ and $\sigma_v$ relax to time-independent $T/\lambda$ and $T/m$, respectively.
Using $\tau_x$ in Eq.~(\ref{relaxation time of x}) and $\tau_v$ in Eq.~(\ref{relaxation time of v}), we can rewrite Eq.~(\ref{omega star}) as
\begin{equation}
  \label{omega star with relaxation times}
  \omega^{*} = \frac{1}{\tau_{v}}\sqrt{1-4\frac{\tau_{v}}{\tau_{x}}}.
\end{equation}
Then, the exponential functions in Eqs.~(\ref{sigma_x with constant lambda})
and (\ref{sigma_v with constant lambda}) are represented by
\begin{align}
  \label{exponential term 1}
  e^{-\frac{\gamma}{m} (t-t_{i})}
  &= e^{-\left(t-t_{i}\right)/\tau_{v}},\\
  \label{exponential term 2}
  e^{-\left(\frac{\gamma}{m} - \omega^{*}\right)(t-t_{i})}
  &= e^{-\left(1-\sqrt{1-4\frac{\tau_{v}}{\tau_{x}}}\right)
    \left(t-t_{i}\right)/\tau_{v}},\\
  \label{exponential term 3}
  e^{-\left(\frac{\gamma}{m} + \omega^{*}\right)(t-t_{i})}
  &= e^{-\left(1+\sqrt{1-4\frac{\tau_{v}}{\tau_{x}}}\right)
    \left(t-t_{i}\right)/\tau_{v}}.
\end{align}

By considering the magnitude relationship between $\tau_x$ and $\tau_v$, we show that the relaxation time of the system is evaluated as $\max(\tau_x,\tau_v)$.
When $\tau_{x} \le 4\tau_{v}$,
$\omega^{*}$ in Eq.~(\ref{omega star with relaxation times}) becomes purely imaginary.
Thus, we can consider that the exponential terms in Eqs.~(\ref{sigma_x with constant lambda}) and (\ref{sigma_v with constant lambda}), which are expressed by Eqs.~(\ref{exponential term 1})--(\ref{exponential term 3}), are sufficiently smaller than the first terms of Eqs.~(\ref{sigma_x with constant lambda}) and (\ref{sigma_v with constant lambda}) when
\begin{equation}
\label{compare the time and relaxation times}
    t-t_{i}\gg \tau_{v}
\end{equation}
is satisfied.
Therefore, we can regard the relaxation time of the system as $\tau_v$.
On the other hand, the case of $\tau_{x} > 4\tau_{v}$ is as follows.
Since the exponential function of the second terms in Eqs.~(\ref{sigma_x with constant lambda}) and (\ref{sigma_v with constant lambda}) is expressed by the relaxation times as in Eq.~(\ref{exponential term 1}), it becomes sufficiently smaller than the first terms when Eq.~(\ref{compare the time and relaxation times}) is satisfied.
Moreover, because the fourth terms in Eqs.~(\ref{sigma_x with constant lambda}) and (\ref{sigma_v with constant lambda}) are expressed by Eq.~(\ref{exponential term 3}) and $1+\sqrt{1-4\tau_v/\tau_x} > 1$, those terms are also sufficiently smaller than the first terms when Eq.~(\ref{compare the time and relaxation times}) is satisfied.
When $4\tau_v/\tau_x$ becomes small, however, $1-\sqrt{1-4\tau_v/\tau_x}$ in the exponent of Eq.~(\ref{exponential term 2}), by which the third terms in Eqs.~(\ref{sigma_x with constant lambda}) and (\ref{sigma_v with constant lambda}) are expressed, approaches 0 and we have to reconsider the case.
When $\tau_x\gg \tau_v$, the exponent of Eq.~(\ref{exponential term 2}) is approximated by
\begin{equation}
  -\frac{1}{\tau_{v}}\left(1-\sqrt{1-4\frac{\tau_{v}}{\tau_{x}}}\right)
    \left(t-t_{i}\right)
  \simeq -\frac{2}{\tau_{x}}\left(t-t_{i}\right),
\end{equation}
which makes Eq.~(\ref{exponential term 2}) vanish when
\begin{equation}
\label{compare the time and relaxation times x}
    t-t_{i}\gg \tau_{x}
\end{equation}
is satisfied. Then, the third terms of Eqs.~(\ref{sigma_x with constant lambda}) and (\ref{sigma_v with constant lambda}) are sufficiently smaller than the first terms.
When $\tau_x\gg \tau_v$, the time for the third terms in Eqs.~(\ref{sigma_x with constant lambda}) and (\ref{sigma_v with constant lambda}) to vanish is longer than the time for the second and fourth terms to vanish.
Therefore, the relaxation time of the system is evaluated as $\tau_x$.
In summary, the relaxation time of the system is represented by
\begin{equation}
\label{relaxation time in small relaxation-times regime}
    \tau\equiv \max(\tau_x,\tau_v).
\end{equation}
Therefore, we can see that when $t-t_{i}\gg \tau$ is satisfied, $\sigma_{x}$ and $\sigma_{v}$ are approximated by
\begin{equation}
  \label{sigma_x with small relaxation times}
  \sigma_{x}\simeq\frac{T}{\lambda},
    \quad
  \sigma_{v} \simeq\frac{T}{m}
\end{equation}
from Eqs.~(\ref{sigma_x with constant lambda}) and (\ref{sigma_v with constant lambda}).
When $\sigma_x$ and $\sigma_v$ are changing toward Eq.~(\ref{sigma_x with small relaxation times}), we consider that the system is in the relaxation.
Moreover, when $t-t_{i}\gg \tau$ is satisfied, we use the phrase ``after the relaxation."
Since the initial condition is included only in $D_1$, $D_2$, and $D_3$ in Eqs.~(\ref{D1})--(\ref{D3}), we find that the variables $\sigma_x$ and $\sigma_v$ relax to the values determined by $T$, $m$, and $\lambda$ even when we choose other initial conditions.
In the limit of $\tau\to 0$, $\sigma_x$ and $\sigma_v$ satisfy Eq.~(\ref{sigma_x with small relaxation times}) when $t-t_i>0$.

From Eq.~(\ref{equation of sigma_x}), we can obtain $\sigma_{xv}$ by differentiating Eq.~(\ref{sigma_x with constant lambda}) with respect to time. 
Because $T$ and $\lambda$ are assumed to be constant, the time derivative of the first term in Eq.~(\ref{sigma_x with constant lambda}) vanishes. 
After the relaxation, we can see that the time derivative of the remaining terms in Eq.~(\ref{sigma_x with constant lambda}) also vanishes. 
Thus, $\sigma_{xv}$ vanishes after the relaxation.

Subsequently, we consider the thermodynamic process where the temperature $T(t)$ and stiffness $\lambda(t)$ depend on time.
In the statement above Eq.~(\ref{ratio of the relaxation times of x}), we assumed that $T(t)$ and $\lambda(t)$ vary smoothly and slowly.
Then, in the small relaxation-times regime, we can expect that the fast relaxation dynamics rapidly vanishes and only the slow dynamics remains accompanying the change of $T(t)$ and $\lambda(t)$.
Therefore, as the resulting approximate dynamics, we obtain the same
expression as Eq.~(\ref{sigma_x with small relaxation times}), by replacing the constant $T$ and $\lambda$ with the time-dependent variables $T(t)$ and $\lambda(t)$, respectively.
Then, we obtain the time derivative of $\sigma_{x}(t)$ and $\sigma_{v}(t)$ after the relaxation in the process as
\begin{equation}
  \label{time dependence of sigma}
  \dot{\sigma}_{x}(t)\simeq \frac{d}{dt}\left(\frac{T(t)}{\lambda(t)}\right)= \frac{T(t)}{\lambda(t)}\left(\frac{d}{dt}\ln \frac{T(t)}{\lambda(t)}\right),
  \
  \dot{\sigma}_{v}(t) \simeq \frac{\dot{T}(t)}{m}.
\end{equation}
From Eqs.~(\ref{equation of sigma_x}) and (\ref{time dependence of sigma}), $\sigma_{xv}(t)$ becomes 
\begin{equation}
  \label{sigma_xv with small relaxation times}
  \sigma_{xv}(t)=\frac{1}{2}\dot{\sigma}_x(t)\simeq \frac{1}{2}\frac{d}{dt}\left(\frac{T(t)}{\lambda(t)}\right)=\frac{T(t)}{2\lambda(t)}\left(\frac{d}{dt}\ln \frac{T(t)}{\lambda(t)}\right).
\end{equation}
Therefore, we obtain the results in Eqs.~(\ref{approximation of variables}) and (\ref{sigma xv in the small relaxation times regime}) in the small relaxation-times regime.
In Appendix~\ref{entropy production at the switching}, we mention that we may need to reconsider these results at the switching between the processes.

\section{DERIVATION OF THE PROTOCOL IN THE FINITE-TIME ADIABATIC PROCESS}
\label{Derivation of the protocol in the adiabatic process}
We derive the protocol of the finite-time adiabatic process by giving a finite $\Delta t$ and the time evolution of $\sigma_x(t)$.
We assume that the finite-time adiabatic process lasts for $t_i\leq t\leq t_f$, and the temperature changes from $T_i$ to $T_f$.
In the finite-time adiabatic process, we need to specify the time evolution of the five variables [$T(t)$ and $\lambda(t)$ in the protocol and $\sigma_x$, $\sigma_v$, and $\sigma_{xv}$ representing the state of the Brownian particle].
In the finite-time adiabatic process in Sec.~\ref{finite-time adiabatic process}, however, there are only four equations in Eqs.~(\ref{equation of sigma_x})--(\ref{equation of sigma_xv}) and (\ref{adiabatic condition}).
Therefore, we have insufficient number of the equations.
However, we can obtain the closed equations if we give the time evolution of one of the five variables.

As we show below, we obtain the protocol by giving the time evolution of $\sigma_x(t)$.
To determine the time evolution of $\sigma_v$, $\sigma_{xv}$, $T$, and $\lambda$, we solve Eqs.~(\ref{equation of sigma_x})--(\ref{equation of sigma_xv}) and (\ref{adiabatic condition}).
From the given $\sigma_x$ and Eq.~(\ref{equation of sigma_x}), we obtain $\sigma_{xv}$.
Using Eqs.~(\ref{equation of sigma_x}) and (\ref{adiabatic condition}), we can rewrite Eqs.~(\ref{equation of sigma_v}) and (\ref{equation of sigma_xv}) as
\begin{align}
  \label{rewrite eq v}
    \dot{T}=&-\lambda\dot{\sigma}_x,\\
    \label{rewrite eq xv}
    m\ddot{\sigma}_{x}=&2T-2\lambda\sigma_x-\gamma\dot{\sigma}_{x}.
\end{align}
From Eq.~(\ref{rewrite eq v}), we obtain $\lambda(t)$ as
\begin{equation}
\label{stiffness in adiabatic process}
    \lambda(t)=-\frac{\dot{T}(t)}{\dot{\sigma}_x(t)}.
\end{equation}
Substituting this into Eq.~(\ref{rewrite eq xv}), we derive the differential equation of $T$ as
\begin{align}
\label{diff eq for T}
    \dot{T}+\left(\frac{d}{dt}\ln\sigma_x\right) T=&\frac{1}{2}\left(\frac{d}{dt}\ln\sigma_x\right)\left(m\ddot{\sigma}_x+\gamma\dot{\sigma}_x\right).
\end{align}
By solving Eq.~(\ref{diff eq for T}), we can derive the time evolution of $T(t)$ as
\begin{align}
\label{T in adiabatic process}
    T(t)=&\frac{1}{2\sigma_x(t)}\left(\gamma\int^{t}_{t_i}\dot{\sigma}_x(t)^2dt\right.\nonumber\\
    &\left.+2T_i\sigma_{xi}+\frac{m}{2}\dot{\sigma}_{x}(t)^2-\frac{m}{2}\dot{\sigma}_{xi}{}^2\right),
\end{align}
using the initial condition $T(t_i)=T_i$.
Then, from Eq.~(\ref{adiabatic condition}), we obtain $\sigma_v$.
Note that $\Delta \Phi$ in the finite-time adiabatic process satisfies 
\begin{align}
\label{Phi in adiabatic process}
\Delta \Phi=& \frac{\sigma_{xf}T_f}{m}-\frac{\sigma_{xi}T_i}{m}-\frac{1}{4}(\dot{\sigma}_{xf}{}^2-\dot{\sigma}_{xi}{}^2)\nonumber\\
=&\frac{1}{2}\frac{\gamma}{m}\int^{t_f}_{t_i} \dot{\sigma}_x(t)^2dt,
\end{align}
where we used Eqs.~(\ref{Phi}), (\ref{equation of sigma_x}), (\ref{adiabatic condition}), and (\ref{Delta Phi}).
From Eqs.~(\ref{T in adiabatic process}) and (\ref{Phi in adiabatic process}), we can confirm that $T(t)$ satisfies $T(t_f)=T_f$.
Substituting the given $\sigma_x(t)$ and $T$ in Eq.~(\ref{T in adiabatic process}) into Eq.~(\ref{stiffness in adiabatic process}), we can obtain the time evolution of $\lambda(t)$.

\section{ENTROPY PRODUCTION AT THE SWITCHINGS}
\label{entropy production at the switching}
We show that the entropy production at the switchings between the isothermal and finite-time adiabatic processes can be neglected in the  small relaxation-times regime.
In Sec.~\ref{Carnot cycle in the small relaxation-times regime}, although we assumed that $T$ and $\lambda$ in our Carnot cycle are continuous at the switchings, we do not assume that the time derivatives of $T$ and $\lambda$ are continuous.
If $\sigma_{xv}$ always satisfies Eq.~(\ref{approximation of variables}) even in the vicinity of the switchings and the time derivative of $T$ and $\lambda$ are discontinuous at the switchings, $\sigma_{xv}$ becomes discontinuous.
However, the variables $\sigma_{x}$, $\sigma_{v}$, and $\sigma_{xv}$ should be continuous because their time evolution is described by the differential equations in Eqs.~(\ref{equation of sigma_x})--(\ref{equation of sigma_xv}).
Thus, we may consider that the variables do not satisfy Eq.~(\ref{approximation of variables}) just after the switchings and relax to Eq.~(\ref{approximation of variables}).
This means that there exists a relaxation just after the switchings.

\begin{figure}[t]
  \includegraphics[width=7.0cm]{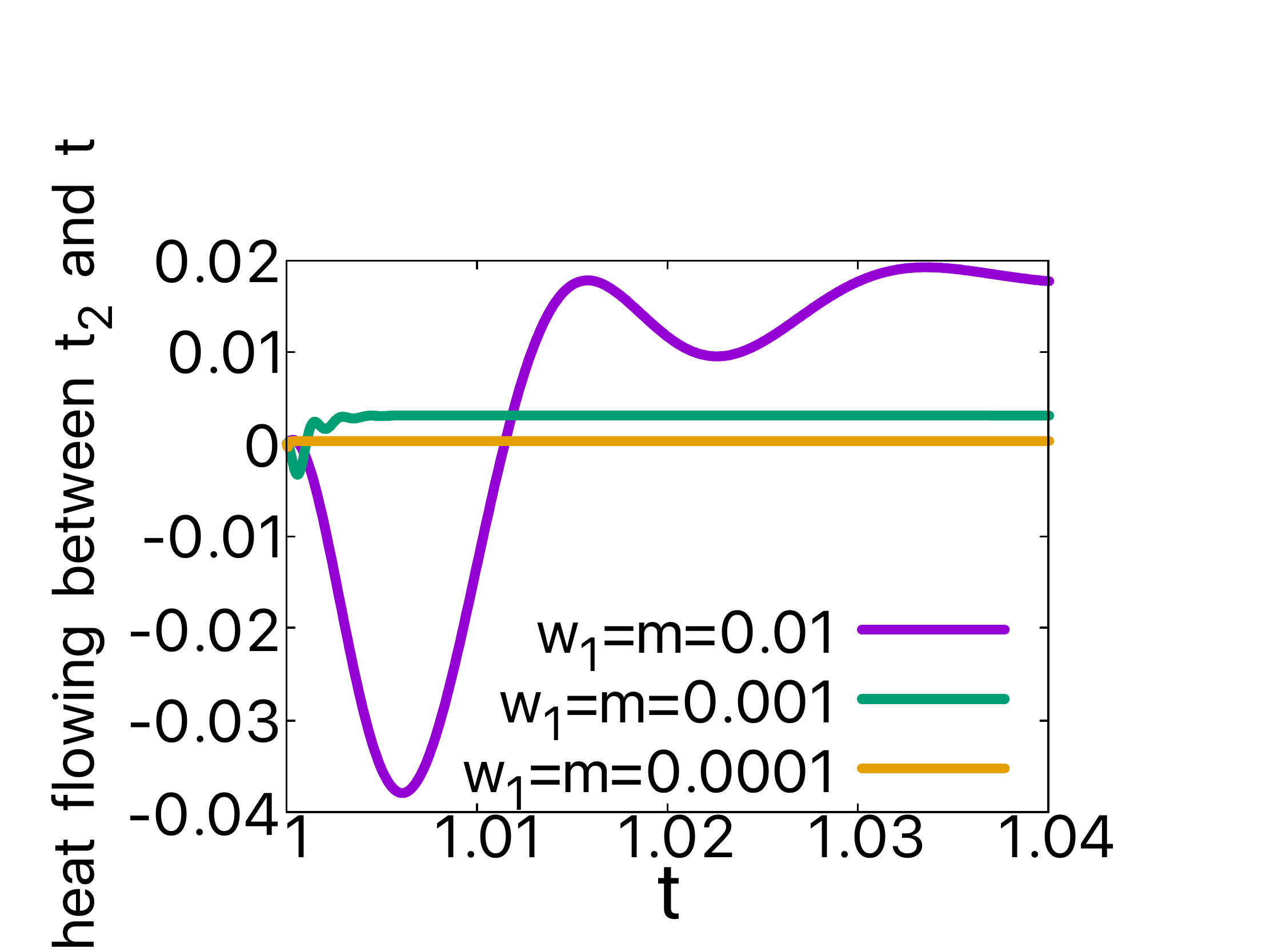}
  \caption{Heat flowing between $t_2$ and $t$ in the finite-time adiabatic process (ii) in Fig.~\ref{fig:schematic_illustration_2021}.
  In this simulation, we used the protocol in Eqs.~(\ref{protocol in the overdamped regime}), (\ref{optimized temperature}), and (\ref{optimized stiffness}) in Sec.~\ref{numerical simulation}.
  We chose $T_{h}=10.0$, $T_{c}=1.0$, and the other parameters as in the numerical simulation in Sec.~\ref{numerical simulation}.
  We here set $t_2=1.0$.
  Thus, the finite-time adiabatic process (ii) lasts for $1.0\leq t\leq 2.0$.  
  We find that although the relaxation after the switching at $t=t_2$ exists, the heat flowing in the relaxation becomes smaller when the relaxation times become smaller.
  }
  \label{fig:rel_heat_local}
\end{figure}
From Eq.~(\ref{Sigma with efficiency}), the entropy production per cycle should vanish to achieve the Carnot efficiency.
Thus, the entropy production in the relaxation after the switchings may affect the efficiency.
We evaluate the entropy production in the small relaxation-times regime and show that it can be neglected.
Since the variables $\sigma_x$, $\sigma_v$, and $\sigma_{xv}$ may not satisfy Eqs.~(\ref{approximation of variables}) and (\ref{sigma xv in the small relaxation times regime}) in the relaxation, we cannot rewrite the entropy production rate by using the relaxation times as shown in Eq.~(\ref{entropy production rate in the small relaxation-times regime}). 
However, since the variables just before the switchings and after the relaxation satisfy Eqs.~(\ref{approximation of variables}) and (\ref{sigma xv in the small relaxation times regime}), we can evaluate the entropy change of the particle and heat flowing in the relaxation as shown below.
Then, by using Eq.~(\ref{def of the entropy production}), we evaluate the entropy production.

Although we here focus on the switching from the hot isothermal process to the finite-time adiabatic process, corresponding to $t=t_2$ in Fig.~\ref{fig:schematic_illustration_2021}, the similar discussion is available in the other switchings.
At that switching, the temperature and stiffness satisfy $T=T_h$ and $\lambda=\lambda_2$, respectively.
When $T$ and $\lambda$ vary smoothly and slowly, as in the statement above Eq.~(\ref{ratio of the relaxation times of x}), we can expect that $T$ and $\lambda$ remain unchanged in the relaxation.
Thus, the entropy change of the particle in this relaxation satisfies
\begin{align}
\label{entropy before the relaxation}
    \Delta S^{\rm rel}\simeq 0
\end{align}
in the small relaxation-times regime because of Eqs.~(\ref{def of the entropy}) and (\ref{approximation of Phi}), where the index ``rel" means the quantity in this relaxation.
Moreover, since $T$ and $\lambda$ are regarded as unchanged in the relaxation, we can see from Eq.~(\ref{approximation of variables}) that each of $\sigma_x$ and $\sigma_v$ just before the relaxation takes the same value as that after the relaxation. 
Therefore, we can evaluate the heat $Q^{\rm rel}$ flowing in this relaxation in the small relaxation-times regime as
\begin{align}
\label{heat in the relaxation}
    Q^{\rm rel}
    \simeq& \frac{1}{2}m(\Delta \sigma_v)^{\rm rel} +\frac{1}{2}\lambda_2(\Delta \sigma_x)^{\rm rel}\simeq 0
\end{align}
from Eqs.~(\ref{def of the heat flux}) and (\ref{def of the heat}).
Figure~\ref{fig:rel_heat_local} shows that the relaxation exists after the switching at $t=t_2$ in the finite-time adiabatic process (ii) in Fig.~\ref{fig:schematic_illustration_2021} with the protocol in Sec.~\ref{numerical simulation}.
However, we can see that the heat flowing in the relaxation becomes smaller when the relaxation times become smaller.
Thus, we can neglect $Q^{\rm rel}$ in Eq.~(\ref{heat in the relaxation}) in the small relaxation-times regime.
Since we can regard $T$ as $T_h$ in the relaxation, we derive the entropy production $\Sigma^{\rm rel}$ of the total system in this relaxation by using Eqs.~(\ref{def of the entropy production}), (\ref{entropy before the relaxation}), and (\ref{heat in the relaxation}) as
\begin{align}
   \Sigma^{\rm rel} \simeq& \Delta S^{\rm rel}-\frac{Q^{\rm rel}}{T_h}
   \simeq 0.
\end{align}
Thus, the entropy production $\Sigma^{\rm rel}$ can be neglected in the small relaxation-times regime.

\bibliography{ref_of_article2021}

\begin{thebibliography}{32}%
\makeatletter
\providecommand \@ifxundefined [1]{%
 \@ifx{#1\undefined}
}%
\providecommand \@ifnum [1]{%
 \ifnum #1\expandafter \@firstoftwo
 \else \expandafter \@secondoftwo
 \fi
}%
\providecommand \@ifx [1]{%
 \ifx #1\expandafter \@firstoftwo
 \else \expandafter \@secondoftwo
 \fi
}%
\providecommand \natexlab [1]{#1}%
\providecommand \enquote  [1]{``#1''}%
\providecommand \bibnamefont  [1]{#1}%
\providecommand \bibfnamefont [1]{#1}%
\providecommand \citenamefont [1]{#1}%
\providecommand \href@noop [0]{\@secondoftwo}%
\providecommand \href [0]{\begingroup \@sanitize@url \@href}%
\providecommand \@href[1]{\@@startlink{#1}\@@href}%
\providecommand \@@href[1]{\endgroup#1\@@endlink}%
\providecommand \@sanitize@url [0]{\catcode `\\12\catcode `\$12\catcode
  `\&12\catcode `\#12\catcode `\^12\catcode `\_12\catcode `\%12\relax}%
\providecommand \@@startlink[1]{}%
\providecommand \@@endlink[0]{}%
\providecommand \url  [0]{\begingroup\@sanitize@url \@url }%
\providecommand \@url [1]{\endgroup\@href {#1}{\urlprefix }}%
\providecommand \urlprefix  [0]{URL }%
\providecommand \Eprint [0]{\href }%
\providecommand \doibase [0]{https://doi.org/}%
\providecommand \selectlanguage [0]{\@gobble}%
\providecommand \bibinfo  [0]{\@secondoftwo}%
\providecommand \bibfield  [0]{\@secondoftwo}%
\providecommand \translation [1]{[#1]}%
\providecommand \BibitemOpen [0]{}%
\providecommand \bibitemStop [0]{}%
\providecommand \bibitemNoStop [0]{.\EOS\space}%
\providecommand \EOS [0]{\spacefactor3000\relax}%
\providecommand \BibitemShut  [1]{\csname bibitem#1\endcsname}%
\let\auto@bib@innerbib\@empty
\bibitem [{\citenamefont {Carnot}(1824)}]{Carnot_book}%
  \BibitemOpen
  \bibfield  {author} {\bibinfo {author} {\bibfnamefont {S.}~\bibnamefont
  {Carnot}},\ }\href@noop {} {\emph {\bibinfo {title} {Reflections on the
  Motive Power of Fire and on Machines Fitted to Develop that Power}}}\
  (\bibinfo  {publisher} {Bachelier, Paris},\ \bibinfo {year}
  {1824})\BibitemShut {NoStop}%
\bibitem [{\citenamefont {Callen}(1985)}]{Callen_2nd_book}%
  \BibitemOpen
  \bibfield  {author} {\bibinfo {author} {\bibfnamefont {H.~B.}\ \bibnamefont
  {Callen}},\ }\href@noop {} {\emph {\bibinfo {title} {Thermodynamics and an
  Introduction to Thermostatistics}}},\ \bibinfo {edition} {2nd}\ ed.\
  (\bibinfo  {publisher} {Wiley, New York},\ \bibinfo {year}
  {1985})\BibitemShut {NoStop}%
\bibitem [{\citenamefont {Benenti}\ \emph {et~al.}(2011)\citenamefont
  {Benenti}, \citenamefont {Saito},\ and\ \citenamefont
  {Casati}}]{PhysRevLett.106.230602}%
  \BibitemOpen
  \bibfield  {author} {\bibinfo {author} {\bibfnamefont {G.}~\bibnamefont
  {Benenti}}, \bibinfo {author} {\bibfnamefont {K.}~\bibnamefont {Saito}},\
  and\ \bibinfo {author} {\bibfnamefont {G.}~\bibnamefont {Casati}},\ }\href
  {https://doi.org/10.1103/PhysRevLett.106.230602} {\bibfield  {journal}
  {\bibinfo  {journal} {Phys. Rev. Lett.}\ }\textbf {\bibinfo {volume} {106}},\
  \bibinfo {pages} {230602} (\bibinfo {year} {2011})}\BibitemShut {NoStop}%
\bibitem [{\citenamefont {Polettini}\ and\ \citenamefont
  {Esposito}(2017)}]{Polettini_2017}%
  \BibitemOpen
  \bibfield  {author} {\bibinfo {author} {\bibfnamefont {M.}~\bibnamefont
  {Polettini}}\ and\ \bibinfo {author} {\bibfnamefont {M.}~\bibnamefont
  {Esposito}},\ }\href {https://doi.org/10.1209/0295-5075/118/40003} {\bibfield
   {journal} {\bibinfo  {journal} {Europhys. Lett.}\ }\textbf {\bibinfo
  {volume} {118}},\ \bibinfo {pages} {40003} (\bibinfo {year}
  {2017})}\BibitemShut {NoStop}%
\bibitem [{\citenamefont {Hondou}\ and\ \citenamefont
  {Sekimoto}(2000)}]{PhysRevE.62.6021}%
  \BibitemOpen
  \bibfield  {author} {\bibinfo {author} {\bibfnamefont {T.}~\bibnamefont
  {Hondou}}\ and\ \bibinfo {author} {\bibfnamefont {K.}~\bibnamefont
  {Sekimoto}},\ }\href {https://doi.org/10.1103/PhysRevE.62.6021} {\bibfield
  {journal} {\bibinfo  {journal} {Phys. Rev. E}\ }\textbf {\bibinfo {volume}
  {62}},\ \bibinfo {pages} {6021} (\bibinfo {year} {2000})}\BibitemShut
  {NoStop}%
\bibitem [{\citenamefont {Shiraishi}(2017)}]{PhysRevE.95.052128}%
  \BibitemOpen
  \bibfield  {author} {\bibinfo {author} {\bibfnamefont {N.}~\bibnamefont
  {Shiraishi}},\ }\href {https://doi.org/10.1103/PhysRevE.95.052128} {\bibfield
   {journal} {\bibinfo  {journal} {Phys. Rev. E}\ }\textbf {\bibinfo {volume}
  {95}},\ \bibinfo {pages} {052128} (\bibinfo {year} {2017})}\BibitemShut
  {NoStop}%
\bibitem [{\citenamefont {Campisi}\ and\ \citenamefont
  {Fazio}(2016)}]{Campisi2016}%
  \BibitemOpen
  \bibfield  {author} {\bibinfo {author} {\bibfnamefont {M.}~\bibnamefont
  {Campisi}}\ and\ \bibinfo {author} {\bibfnamefont {R.}~\bibnamefont
  {Fazio}},\ }\href {https://doi.org/10.1038/ncomms11895} {\bibfield  {journal}
  {\bibinfo  {journal} {Nat. Comm.}\ }\textbf {\bibinfo {volume} {7}},\
  \bibinfo {pages} {11895} (\bibinfo {year} {2016})}\BibitemShut {NoStop}%
\bibitem [{\citenamefont {Brandner}\ \emph {et~al.}(2013)\citenamefont
  {Brandner}, \citenamefont {Saito},\ and\ \citenamefont
  {Seifert}}]{PhysRevLett.110.070603}%
  \BibitemOpen
  \bibfield  {author} {\bibinfo {author} {\bibfnamefont {K.}~\bibnamefont
  {Brandner}}, \bibinfo {author} {\bibfnamefont {K.}~\bibnamefont {Saito}},\
  and\ \bibinfo {author} {\bibfnamefont {U.}~\bibnamefont {Seifert}},\ }\href
  {https://doi.org/10.1103/PhysRevLett.110.070603} {\bibfield  {journal}
  {\bibinfo  {journal} {Phys. Rev. Lett.}\ }\textbf {\bibinfo {volume} {110}},\
  \bibinfo {pages} {070603} (\bibinfo {year} {2013})}\BibitemShut {NoStop}%
\bibitem [{\citenamefont {Balachandran}\ \emph {et~al.}(2013)\citenamefont
  {Balachandran}, \citenamefont {Benenti},\ and\ \citenamefont
  {Casati}}]{PhysRevB.87.165419}%
  \BibitemOpen
  \bibfield  {author} {\bibinfo {author} {\bibfnamefont {V.}~\bibnamefont
  {Balachandran}}, \bibinfo {author} {\bibfnamefont {G.}~\bibnamefont
  {Benenti}},\ and\ \bibinfo {author} {\bibfnamefont {G.}~\bibnamefont
  {Casati}},\ }\href {https://doi.org/10.1103/PhysRevB.87.165419} {\bibfield
  {journal} {\bibinfo  {journal} {Phys. Rev. B}\ }\textbf {\bibinfo {volume}
  {87}},\ \bibinfo {pages} {165419} (\bibinfo {year} {2013})}\BibitemShut
  {NoStop}%
\bibitem [{\citenamefont {Yamamoto}\ \emph {et~al.}(2016)\citenamefont
  {Yamamoto}, \citenamefont {Entin-Wohlman}, \citenamefont {Aharony},\ and\
  \citenamefont {Hatano}}]{PhysRevB.94.121402}%
  \BibitemOpen
  \bibfield  {author} {\bibinfo {author} {\bibfnamefont {K.}~\bibnamefont
  {Yamamoto}}, \bibinfo {author} {\bibfnamefont {O.}~\bibnamefont
  {Entin-Wohlman}}, \bibinfo {author} {\bibfnamefont {A.}~\bibnamefont
  {Aharony}},\ and\ \bibinfo {author} {\bibfnamefont {N.}~\bibnamefont
  {Hatano}},\ }\href {https://doi.org/10.1103/PhysRevB.94.121402} {\bibfield
  {journal} {\bibinfo  {journal} {Phys. Rev. B}\ }\textbf {\bibinfo {volume}
  {94}},\ \bibinfo {pages} {121402(R)} (\bibinfo {year} {2016})}\BibitemShut
  {NoStop}%
\bibitem [{\citenamefont {Stark}\ \emph {et~al.}(2014)\citenamefont {Stark},
  \citenamefont {Brandner}, \citenamefont {Saito},\ and\ \citenamefont
  {Seifert}}]{PhysRevLett.112.140601}%
  \BibitemOpen
  \bibfield  {author} {\bibinfo {author} {\bibfnamefont {J.}~\bibnamefont
  {Stark}}, \bibinfo {author} {\bibfnamefont {K.}~\bibnamefont {Brandner}},
  \bibinfo {author} {\bibfnamefont {K.}~\bibnamefont {Saito}},\ and\ \bibinfo
  {author} {\bibfnamefont {U.}~\bibnamefont {Seifert}},\ }\href
  {https://doi.org/10.1103/PhysRevLett.112.140601} {\bibfield  {journal}
  {\bibinfo  {journal} {Phys. Rev. Lett.}\ }\textbf {\bibinfo {volume} {112}},\
  \bibinfo {pages} {140601} (\bibinfo {year} {2014})}\BibitemShut {NoStop}%
\bibitem [{\citenamefont {S\'anchez}\ \emph {et~al.}(2015)\citenamefont
  {S\'anchez}, \citenamefont {Sothmann},\ and\ \citenamefont
  {Jordan}}]{PhysRevLett.114.146801}%
  \BibitemOpen
  \bibfield  {author} {\bibinfo {author} {\bibfnamefont {R.}~\bibnamefont
  {S\'anchez}}, \bibinfo {author} {\bibfnamefont {B.}~\bibnamefont
  {Sothmann}},\ and\ \bibinfo {author} {\bibfnamefont {A.~N.}\ \bibnamefont
  {Jordan}},\ }\href {https://doi.org/10.1103/PhysRevLett.114.146801}
  {\bibfield  {journal} {\bibinfo  {journal} {Phys. Rev. Lett.}\ }\textbf
  {\bibinfo {volume} {114}},\ \bibinfo {pages} {146801} (\bibinfo {year}
  {2015})}\BibitemShut {NoStop}%
\bibitem [{\citenamefont {Sothmann}\ \emph {et~al.}(2014)\citenamefont
  {Sothmann}, \citenamefont {S{\'{a}}nchez},\ and\ \citenamefont
  {Jordan}}]{Sothmann_2014}%
  \BibitemOpen
  \bibfield  {author} {\bibinfo {author} {\bibfnamefont {B.}~\bibnamefont
  {Sothmann}}, \bibinfo {author} {\bibfnamefont {R.}~\bibnamefont
  {S{\'{a}}nchez}},\ and\ \bibinfo {author} {\bibfnamefont {A.~N.}\
  \bibnamefont {Jordan}},\ }\href {https://doi.org/10.1209/0295-5075/107/47003}
  {\bibfield  {journal} {\bibinfo  {journal} {Europhys. Lett}\ }\textbf
  {\bibinfo {volume} {107}},\ \bibinfo {pages} {47003} (\bibinfo {year}
  {2014})}\BibitemShut {NoStop}%
\bibitem [{\citenamefont {Ma}\ \emph {et~al.}(2018)\citenamefont {Ma},
  \citenamefont {Xu}, \citenamefont {Dong},\ and\ \citenamefont
  {Sun}}]{PhysRevE.98.042112}%
  \BibitemOpen
  \bibfield  {author} {\bibinfo {author} {\bibfnamefont {Y.-H.}\ \bibnamefont
  {Ma}}, \bibinfo {author} {\bibfnamefont {D.}~\bibnamefont {Xu}}, \bibinfo
  {author} {\bibfnamefont {H.}~\bibnamefont {Dong}},\ and\ \bibinfo {author}
  {\bibfnamefont {C.-P.}\ \bibnamefont {Sun}},\ }\href
  {https://doi.org/10.1103/PhysRevE.98.042112} {\bibfield  {journal} {\bibinfo
  {journal} {Phys. Rev. E}\ }\textbf {\bibinfo {volume} {98}},\ \bibinfo
  {pages} {042112} (\bibinfo {year} {2018})}\BibitemShut {NoStop}%
\bibitem [{\citenamefont {Abiuso}\ and\ \citenamefont
  {Perarnau-Llobet}(2020)}]{PhysRevLett.124.110606}%
  \BibitemOpen
  \bibfield  {author} {\bibinfo {author} {\bibfnamefont {P.}~\bibnamefont
  {Abiuso}}\ and\ \bibinfo {author} {\bibfnamefont {M.}~\bibnamefont
  {Perarnau-Llobet}},\ }\href {https://doi.org/10.1103/PhysRevLett.124.110606}
  {\bibfield  {journal} {\bibinfo  {journal} {Phys. Rev. Lett.}\ }\textbf
  {\bibinfo {volume} {124}},\ \bibinfo {pages} {110606} (\bibinfo {year}
  {2020})}\BibitemShut {NoStop}%
\bibitem [{\citenamefont {Holubec}\ and\ \citenamefont
  {Ryabov}(2017)}]{PhysRevE.96.062107}%
  \BibitemOpen
  \bibfield  {author} {\bibinfo {author} {\bibfnamefont {V.}~\bibnamefont
  {Holubec}}\ and\ \bibinfo {author} {\bibfnamefont {A.}~\bibnamefont
  {Ryabov}},\ }\href {https://doi.org/10.1103/PhysRevE.96.062107} {\bibfield
  {journal} {\bibinfo  {journal} {Phys. Rev. E}\ }\textbf {\bibinfo {volume}
  {96}},\ \bibinfo {pages} {062107} (\bibinfo {year} {2017})}\BibitemShut
  {NoStop}%
\bibitem [{\citenamefont {Holubec}\ and\ \citenamefont
  {Ryabov}(2018)}]{PhysRevLett.121.120601}%
  \BibitemOpen
  \bibfield  {author} {\bibinfo {author} {\bibfnamefont {V.}~\bibnamefont
  {Holubec}}\ and\ \bibinfo {author} {\bibfnamefont {A.}~\bibnamefont
  {Ryabov}},\ }\href {https://doi.org/10.1103/PhysRevLett.121.120601}
  {\bibfield  {journal} {\bibinfo  {journal} {Phys. Rev. Lett.}\ }\textbf
  {\bibinfo {volume} {121}},\ \bibinfo {pages} {120601} (\bibinfo {year}
  {2018})}\BibitemShut {NoStop}%
\bibitem [{\citenamefont {Miura}\ \emph {et~al.}(2021)\citenamefont {Miura},
  \citenamefont {Izumida},\ and\ \citenamefont {Okuda}}]{PhysRevE.103.042125}%
  \BibitemOpen
  \bibfield  {author} {\bibinfo {author} {\bibfnamefont {K.}~\bibnamefont
  {Miura}}, \bibinfo {author} {\bibfnamefont {Y.}~\bibnamefont {Izumida}},\
  and\ \bibinfo {author} {\bibfnamefont {K.}~\bibnamefont {Okuda}},\ }\href
  {https://doi.org/10.1103/PhysRevE.103.042125} {\bibfield  {journal} {\bibinfo
   {journal} {Phys. Rev. E}\ }\textbf {\bibinfo {volume} {103}},\ \bibinfo
  {pages} {042125} (\bibinfo {year} {2021})}\BibitemShut {NoStop}%
\bibitem [{\citenamefont {Lee}\ and\ \citenamefont {Park}(2017)}]{Lee_2017}%
  \BibitemOpen
  \bibfield  {author} {\bibinfo {author} {\bibfnamefont {J.~S.}\ \bibnamefont
  {Lee}}\ and\ \bibinfo {author} {\bibfnamefont {H.}~\bibnamefont {Park}},\
  }\href {https://doi.org/10.1038/s41598-017-10664-9} {\bibfield  {journal}
  {\bibinfo  {journal} {Scientific Reports}\ }\textbf {\bibinfo {volume} {7}},\
  \bibinfo {pages} {10725} (\bibinfo {year} {2017})}\BibitemShut {NoStop}%
\bibitem [{\citenamefont {Lee}\ \emph {et~al.}(2020)\citenamefont {Lee},
  \citenamefont {Park}, \citenamefont {Chun}, \citenamefont {Um},\ and\
  \citenamefont {Park}}]{PhysRevE.101.052132}%
  \BibitemOpen
  \bibfield  {author} {\bibinfo {author} {\bibfnamefont {J.~S.}\ \bibnamefont
  {Lee}}, \bibinfo {author} {\bibfnamefont {J.-M.}\ \bibnamefont {Park}},
  \bibinfo {author} {\bibfnamefont {H.-M.}\ \bibnamefont {Chun}}, \bibinfo
  {author} {\bibfnamefont {J.}~\bibnamefont {Um}},\ and\ \bibinfo {author}
  {\bibfnamefont {H.}~\bibnamefont {Park}},\ }\href
  {https://doi.org/10.1103/PhysRevE.101.052132} {\bibfield  {journal} {\bibinfo
   {journal} {Phys. Rev. E}\ }\textbf {\bibinfo {volume} {101}},\ \bibinfo
  {pages} {052132} (\bibinfo {year} {2020})}\BibitemShut {NoStop}%
\bibitem [{\citenamefont {Shiraishi}\ \emph {et~al.}(2016)\citenamefont
  {Shiraishi}, \citenamefont {Saito},\ and\ \citenamefont
  {Tasaki}}]{PhysRevLett.117.190601}%
  \BibitemOpen
  \bibfield  {author} {\bibinfo {author} {\bibfnamefont {N.}~\bibnamefont
  {Shiraishi}}, \bibinfo {author} {\bibfnamefont {K.}~\bibnamefont {Saito}},\
  and\ \bibinfo {author} {\bibfnamefont {H.}~\bibnamefont {Tasaki}},\ }\href
  {https://doi.org/10.1103/PhysRevLett.117.190601} {\bibfield  {journal}
  {\bibinfo  {journal} {Phys. Rev. Lett.}\ }\textbf {\bibinfo {volume} {117}},\
  \bibinfo {pages} {190601} (\bibinfo {year} {2016})}\BibitemShut {NoStop}%
\bibitem [{\citenamefont {Pietzonka}\ and\ \citenamefont
  {Seifert}(2018)}]{PhysRevLett.120.190602}%
  \BibitemOpen
  \bibfield  {author} {\bibinfo {author} {\bibfnamefont {P.}~\bibnamefont
  {Pietzonka}}\ and\ \bibinfo {author} {\bibfnamefont {U.}~\bibnamefont
  {Seifert}},\ }\href {https://doi.org/10.1103/PhysRevLett.120.190602}
  {\bibfield  {journal} {\bibinfo  {journal} {Phys. Rev. Lett.}\ }\textbf
  {\bibinfo {volume} {120}},\ \bibinfo {pages} {190602} (\bibinfo {year}
  {2018})}\BibitemShut {NoStop}%
\bibitem [{\citenamefont {Koyuk}\ \emph {et~al.}(2018)\citenamefont {Koyuk},
  \citenamefont {Seifert},\ and\ \citenamefont {Pietzonka}}]{Koyuk_2018}%
  \BibitemOpen
  \bibfield  {author} {\bibinfo {author} {\bibfnamefont {T.}~\bibnamefont
  {Koyuk}}, \bibinfo {author} {\bibfnamefont {U.}~\bibnamefont {Seifert}},\
  and\ \bibinfo {author} {\bibfnamefont {P.}~\bibnamefont {Pietzonka}},\ }\href
  {https://doi.org/10.1088/1751-8121/aaeec4} {\bibfield  {journal} {\bibinfo
  {journal} {J. Phys. A: Math. Theor.}\ }\textbf {\bibinfo {volume} {52}},\
  \bibinfo {pages} {02LT02} (\bibinfo {year} {2018})}\BibitemShut {NoStop}%
\bibitem [{\citenamefont {Dechant}(2018)}]{Dechant_2018}%
  \BibitemOpen
  \bibfield  {author} {\bibinfo {author} {\bibfnamefont {A.}~\bibnamefont
  {Dechant}},\ }\href {https://doi.org/10.1088/1751-8121/aaf3ff} {\bibfield
  {journal} {\bibinfo  {journal} {Journal of Physics A: Mathematical and
  Theoretical}\ }\textbf {\bibinfo {volume} {52}},\ \bibinfo {pages} {035001}
  (\bibinfo {year} {2018})}\BibitemShut {NoStop}%
\bibitem [{\citenamefont {Dechant}\ and\ \citenamefont
  {Sasa}(2018)}]{PhysRevE.97.062101}%
  \BibitemOpen
  \bibfield  {author} {\bibinfo {author} {\bibfnamefont {A.}~\bibnamefont
  {Dechant}}\ and\ \bibinfo {author} {\bibfnamefont {S.-i.}\ \bibnamefont
  {Sasa}},\ }\href {https://doi.org/10.1103/PhysRevE.97.062101} {\bibfield
  {journal} {\bibinfo  {journal} {Phys. Rev. E}\ }\textbf {\bibinfo {volume}
  {97}},\ \bibinfo {pages} {062101} (\bibinfo {year} {2018})}\BibitemShut
  {NoStop}%
\bibitem [{\citenamefont {Dechant}\ \emph {et~al.}(2017)\citenamefont
  {Dechant}, \citenamefont {Kiesel},\ and\ \citenamefont
  {Lutz}}]{Dechant_2017}%
  \BibitemOpen
  \bibfield  {author} {\bibinfo {author} {\bibfnamefont {A.}~\bibnamefont
  {Dechant}}, \bibinfo {author} {\bibfnamefont {N.}~\bibnamefont {Kiesel}},\
  and\ \bibinfo {author} {\bibfnamefont {E.}~\bibnamefont {Lutz}},\ }\href
  {https://doi.org/10.1209/0295-5075/119/50003} {\bibfield  {journal} {\bibinfo
   {journal} {Europhys. Lett.}\ }\textbf {\bibinfo {volume} {119}},\ \bibinfo
  {pages} {50003} (\bibinfo {year} {2017})}\BibitemShut {NoStop}%
\bibitem [{\citenamefont {Schmiedl}\ and\ \citenamefont
  {Seifert}(2008)}]{Schmiedl_2007}%
  \BibitemOpen
  \bibfield  {author} {\bibinfo {author} {\bibfnamefont {T.}~\bibnamefont
  {Schmiedl}}\ and\ \bibinfo {author} {\bibfnamefont {U.}~\bibnamefont
  {Seifert}},\ }\href {https://doi.org/10.1209/0295-5075/81/20003} {\bibfield
  {journal} {\bibinfo  {journal} {Europhys. Lett.}\ }\textbf {\bibinfo {volume}
  {81}},\ \bibinfo {pages} {20003} (\bibinfo {year} {2008})}\BibitemShut
  {NoStop}%
\bibitem [{\citenamefont {Arold}\ \emph {et~al.}(2018)\citenamefont {Arold},
  \citenamefont {Dechant},\ and\ \citenamefont {Lutz}}]{PhysRevE.97.022131}%
  \BibitemOpen
  \bibfield  {author} {\bibinfo {author} {\bibfnamefont {D.}~\bibnamefont
  {Arold}}, \bibinfo {author} {\bibfnamefont {A.}~\bibnamefont {Dechant}},\
  and\ \bibinfo {author} {\bibfnamefont {E.}~\bibnamefont {Lutz}},\ }\href
  {https://doi.org/10.1103/PhysRevE.97.022131} {\bibfield  {journal} {\bibinfo
  {journal} {Phys. Rev. E}\ }\textbf {\bibinfo {volume} {97}},\ \bibinfo
  {pages} {022131} (\bibinfo {year} {2018})}\BibitemShut {NoStop}%
\bibitem [{\citenamefont {Mart\'{\i}nez}\ \emph {et~al.}(2015)\citenamefont
  {Mart\'{\i}nez}, \citenamefont {Rold\'an}, \citenamefont {Dinis},
  \citenamefont {Petrov},\ and\ \citenamefont {Rica}}]{PhysRevLett.114.120601}%
  \BibitemOpen
  \bibfield  {author} {\bibinfo {author} {\bibfnamefont {I.~A.}\ \bibnamefont
  {Mart\'{\i}nez}}, \bibinfo {author} {\bibfnamefont {E.}~\bibnamefont
  {Rold\'an}}, \bibinfo {author} {\bibfnamefont {L.}~\bibnamefont {Dinis}},
  \bibinfo {author} {\bibfnamefont {D.}~\bibnamefont {Petrov}},\ and\ \bibinfo
  {author} {\bibfnamefont {R.~A.}\ \bibnamefont {Rica}},\ }\href
  {https://doi.org/10.1103/PhysRevLett.114.120601} {\bibfield  {journal}
  {\bibinfo  {journal} {Phys. Rev. Lett.}\ }\textbf {\bibinfo {volume} {114}},\
  \bibinfo {pages} {120601} (\bibinfo {year} {2015})}\BibitemShut {NoStop}%
\bibitem [{\citenamefont {Plata}\ \emph
  {et~al.}(2020{\natexlab{a}})\citenamefont {Plata}, \citenamefont
  {Gu\'ery-Odelin}, \citenamefont {Trizac},\ and\ \citenamefont
  {Prados}}]{PhysRevE.101.032129}%
  \BibitemOpen
  \bibfield  {author} {\bibinfo {author} {\bibfnamefont {C.~A.}\ \bibnamefont
  {Plata}}, \bibinfo {author} {\bibfnamefont {D.}~\bibnamefont
  {Gu\'ery-Odelin}}, \bibinfo {author} {\bibfnamefont {E.}~\bibnamefont
  {Trizac}},\ and\ \bibinfo {author} {\bibfnamefont {A.}~\bibnamefont
  {Prados}},\ }\href {https://doi.org/10.1103/PhysRevE.101.032129} {\bibfield
  {journal} {\bibinfo  {journal} {Phys. Rev. E}\ }\textbf {\bibinfo {volume}
  {101}},\ \bibinfo {pages} {032129} (\bibinfo {year}
  {2020}{\natexlab{a}})}\BibitemShut {NoStop}%
\bibitem [{\citenamefont {Plata}\ \emph
  {et~al.}(2020{\natexlab{b}})\citenamefont {Plata}, \citenamefont
  {Gu{\'{e}}ry-Odelin}, \citenamefont {Trizac},\ and\ \citenamefont
  {Prados}}]{Plata_2020}%
  \BibitemOpen
  \bibfield  {author} {\bibinfo {author} {\bibfnamefont {C.~A.}\ \bibnamefont
  {Plata}}, \bibinfo {author} {\bibfnamefont {D.}~\bibnamefont
  {Gu{\'{e}}ry-Odelin}}, \bibinfo {author} {\bibfnamefont {E.}~\bibnamefont
  {Trizac}},\ and\ \bibinfo {author} {\bibfnamefont {A.}~\bibnamefont
  {Prados}},\ }\href {https://doi.org/10.1088/1742-5468/abb0e1} {\bibfield
  {journal} {\bibinfo  {journal} {J. Stat. Mech.}\ }\textbf {\bibinfo {volume}
  {2020}},\ \bibinfo {pages} {093207} (\bibinfo {year}
  {2020}{\natexlab{b}})}\BibitemShut {NoStop}%
\bibitem [{\citenamefont {Risken}(1996)}]{Risken_2nd_book}%
  \BibitemOpen
  \bibfield  {author} {\bibinfo {author} {\bibfnamefont {H.}~\bibnamefont
  {Risken}},\ }\href@noop {} {\emph {\bibinfo {title} {The Fokker-Planck
  Equation}}},\ \bibinfo {edition} {2nd}\ ed.\ (\bibinfo  {publisher}
  {Springer, New York},\ \bibinfo {year} {1996})\BibitemShut {NoStop}%
\end{thebibliography}%

\end{document}